\documentclass[journal]{IEEEtran}
\usepackage{graphicx}
\usepackage{epstopdf}
\usepackage{fancyhdr}
\usepackage{amsmath,amssymb,amstext}
\usepackage{subeqnarray}
\usepackage{cite}
\usepackage{hhline}
\usepackage{bm}
\usepackage{soul}
\usepackage{times,amsmath,amsfonts}
\usepackage{subfigure}
\usepackage{amsmath,amssymb}
\usepackage{graphicx,picture}
\usepackage{cite}
\usepackage{subfigure}
\usepackage{subeqnarray}
\usepackage{cases}
\usepackage{color}
\usepackage{overpic}
\usepackage{multirow}
\usepackage{booktabs}
\ifCLASSINFOpdf
\else
   \usepackage{graphicx}
   \graphicspath{{../eps/}}
   \DeclareGraphicsExtensions{.eps}
\fi
\usepackage{fancyhdr}
\usepackage{amsmath,amssymb,amstext}
\usepackage{subeqnarray}
\usepackage{cite}
\usepackage{hhline}
\usepackage{bm}
\usepackage{soul}
\usepackage{times,amsmath,amsfonts}
\usepackage{subfigure}
\usepackage{amsmath,amssymb}
\usepackage{graphicx,picture}
\usepackage{cite}
\usepackage{subfigure}
\usepackage{subeqnarray}
\usepackage{cases}
\usepackage{color}
\usepackage{overpic}  % removed: requires epic.sty (unavailable), not used
\usepackage{multirow}
\usepackage{booktabs}
\usepackage{epstopdf}
\usepackage[ruled, linesnumbered]{algorithm2e}

\definecolor{r}{rgb}{1,0,0}
\definecolor{b}{rgb}{0,0,1}
\definecolor{k}{rgb}{0,1,1}
\usepackage{upgreek}

\newcounter{saveeqn}%

\allowdisplaybreaks
\DeclareMathSymbol{\Phi}{\mathord}{letters}{8}

\begin{document}
\title{Low-Altitude ISAC With Spherical Directly-Connected Antenna Array: Performance Analysis and Beamforming Optimization}

 \author{
 \IEEEauthorblockN{Zhiqiang Xiao, Tao Zhang, Zhenjun Dong, Hao Wu, Xiaoqiang Xiao, and Jianhua, Zhang,~\IEEEmembership{Fellow, IEEE}}

  \thanks{
  %This work was supported by the Natural Science Foundation for Distinguished Young
 % %Scholars of Jiangsu Province with grant number BK20240070.
 Z. Xiao, T. Zhang, H. Wu and X. Qiao are with the Sixty-Third Research Institute, National University of Defense Technology, Nanjing, 210007, China (e-mail: zhiqiang\_xiao@nudt.edu.cn, ztcool@126.com).
 %(\emph{Corresponding author: Tao Zhang})

 Z. Dong is with Purple Mountain Laboratories, Nanjing 211111, China (e-mail:zhenjun\_dong@seu.edu.cn).

 J. Zhang is with the State Key Lab of Networking and Switching Technology, Beijing University of Posts and Telecommunications, Beijing 100876, China (e-mail: jhzhang@bupt.edu.cn).

 % Y. Zeng is with the National Mobile Communications Research Laboratory, Southeast University, Nanjing, 210096, China, and also with the Purple Mountain Laboratories, Nanjing, 211111.
 %
 }
 }
\maketitle

\begin{abstract}
The safety development requirements of low-altitude economy (LAE) renders the robust low-altitude airspace monitoring critical important than ever before.
Integrated sensing and communication (ISAC) as one of the key development trends of 6G provides potential solutions for the LAE.
However, conventional antenna arrays suffer from limited three-dimensional (3D) sensing coverage and degraded angular resolution at high elevation angles.
To address these challenges, this paper investigates low-altitude ISAC systems enabled by the recently proposed spherical directly-connected antenna array (DCAA).
By carefully deploying multiple simple uniform planar arrays (sUPAs) over a spherical surface, without relying on any phase shifter, spherical DCAA enjoys advantages of full 3D coverage, superior and uniform angular resolution, enhanced energy-focusing and low hardware cost.
In this paper, we first characterizes the sensing performance of the spherical DCAA, in terms of the sensing signal-to-noise ratio (SNR), area average probability of detection, and Cramér-Rao lower bounds (CRLBs) for both elevation and azimuth angle estimation.
Then, a low-altitude ISAC optimization problem is formulated to maximize the worst-case sensing SNR over a prescribed aerial region while satisfying the communication quality-of-service requirements of ground users.
To effective solve this mixed-integer non-convex problem, we develop a novel greed-based joint array selection and beamforming optimization framework.
Simulation results demonstrate that spherical DCAA significantly outperforms conventional UPA in terms of sensing coverage, angle estimation accuracy, and communication-sensing SNR tradeoff, highlighting its potential for future low-altitude ISAC systems.
\end{abstract}


\section{Introduction}
Low-altitude economy (LAE) has attracted significant attention recent years, driven by various emerging applications, such as aerial express delivery, drone light show, and three-dimensional (3D) environment mapping.
However, due to the complex and dynamic low-altitude environment, aerial platforms including both manned aircrafts and unmanned aerial vehicles (UAV) are exposed to substantial risks of flight accidents, such as collisions, crashes, and airspace intrusions.
This renders the real-time low-altitude airspace sensing critical important for the safety development of the LAE \cite{jiang2025integrated, song2025overview,jun2026low}.

As an important development trend of sixth-generation (6G) wireless networks, integrated sensing and communication (ISAC) holds great potentials for supporting the LAE.
It is envisioned that densely deployed base stations (BSs) will be capable of simultaneously providing reliable communication and pervasive sensing services in the low-altitude environment \cite{cheng2025networked, zhao2025networked, zhang2025unified}.
To this end, extensive research efforts have been devoted to enhancing the sensing capabilities of BSs, including information theory \cite{xiong2023fundamental}, waveform design \cite{xiao2022waveform}, codebook design \cite{xiao2024simultaneous}, network deployment \cite{meng2024cooperative}, and prototyping experiments~\cite{luo2025isac}, etc.
However, most of the existing works are developed based on the conventional planar arrays, which face fundamental challenging in supporting high-performance low-altitude ISAC.

First, due to the high mobility and 3D spatial distribution of aerial targets,
antenna arrays are required to provide full 3D coverage for seamless low-altitude sensing.
However, conventional planar arrays, such as uniform planar array (UPA), are typically designed to provide directional beamforming within a limited angular sector \cite{zhang20183d}.
This may lead to the sensing coverage hole, particularly in the region directly above the BS, that significantly degrades the low-altitude sensing performance.
Although cooperative sensing among multiple BSs can partially alleviate this issue, it inevitably introduces additional signaling overhead and coordination complexity~\cite{wang2024isac,tang2025cooperative,liu2025cooperative}.
Second, the angular resolution of conventional UPA deteriorates considerably as the target direction departs from the array boresight.
Since aerial targets are frequently located in high-elevation region, conventional UPAs often suffer from insufficient angular resolution to distinguish closely spaced aerial targets \cite{min2025integrated}.
Third, although increasing the array aperture can improve both communication and sensing performance, it leads to the rapidly increasing hardware costs.
Even with the hybrid analog/digital beamforming (HBF), which significantly reduces the number of radio frequency (RF) chains, a large number of phase shifters is still required.
Furthermore, implementing high-resolution phase shifter remains challenging particularly at millimeter-wave (mmWave) and Terahertz (THz) frequency bands \cite{el2014spatially,deng2023compact}.

To address these challenges, recent research has explored several advanced antenna array architectures.
For example, lens antenna array can focus the electromagnetic waves onto a small subset of antenna array, thereby significantly reducing the number of RF chains and hardware cost \cite{zeng2018multi}.
However, as the energy-focusing effect can be essentially interpreted as a spatial Fourier transformation that maps signals into the beamspace domain, lens antenna array still suffers from the degraded angular resolution in high elevation region \cite{shim2018cramer}.
With the development of high-speed antenna switching and control technologies, the general concept of reconfigurable antenna array has been recently proposed \cite{castellanos2025embracing, wang2026reconfigurable}.
Unlike conventional arrays with the static architecture, reconfigurable antenna arrays can adaptively reconfigure the array geometries through antenna selection, movement, or rotation, provding additional spatial degrees of freedom (DoFs) for both communication and sensing performance improvement.
Representative examples include sparse multi-input multi-output (MIMO) \cite{li2025sparse}, fluid antenna (FA) \cite{zhou2024fluid}, movable antenna (MA) \cite{ma2024movable}, pinching antenna (PA) \cite{ouyang2026rate}, and rotatable antenna (RA) \cite{xiong2025intelligent}.
Furthermore, the tri-hybrid MIMO architecture has been proposed in \cite{castellanos2025embracing} and \cite{heath2025tri}, which combines conventional HBF with reconfigurable antenna array.

Although the aforementioned reconfigurable antenna arrays provide more design DoFs to enhance both communication and sensing performance, they also incur high hardware cost and control overhead.
Different from above these, a novel architecture termed directly-connected antenna array (DCAA) has been  recently proposed \cite{dong2026novel,jiang2025ray,xiao2026ray,zhu2026cost,jiang20263d}.
The key idea of DCAA is to exploit the inherent energy-focusing capability of simple antenna arrays.
Note that even without phase shifters or digital beamforming, electromagnetic waves can still be naturally concentrated around the array boresight.
Therefore, by properly deploying the orientations of multiple antenna arrays and dynamically selecting the most suitable arrays to connect with the a less number of RF chains, DCAA can achieve flexible beamforming with significant reduced hardware cost.

Building upon the DCAA concept, the ray antenna array (RAA) was first proposed \cite{dong2026novel}, which consists of multiple simple uniform linear arrays (sULAs) arranged in a ray-like structure.
Note that all elements within each sULA are directly connected to a common analog combiner or power splitter, while a switch network is employed for the sULA selection.
By carefully designing the orientations of sULAs, RAA can achieve direction-independent uniform angular resolution and enhanced energy-focusing with low hardware cost.
The authors in \cite{jiang2025ray} and \cite{xiao2026ray} have investigated RAA for low-altitude ISAC, where both analytical and simulation results demonstrate the superiority of RAA over conventional ULA in terms of sensing coverage and angular resolution.
However, RAA faces practical deployment challenges due to the potential blockage between adjacent sULAs.
To mitigate this issue, a cylindrical DCAA was proposed \cite{zhu2026cost}, where multiple sub-arrays are stacked along the vertical direction to avoid mutual blockage, however it is difficult to provide full 3D coverage.
More recently, a spherical DCAA was proposed \cite{jiang20263d}, which generalizes the RAA from a planar to a spherical configuration.
The spherical DCAA consists of multiple simple UPAs (sUPAs) distributed over a spherical surface with optimized orientations.
Note that the spherical DCAA inherits the key advantages of RAA and can further realize full 3D angular coverage.
As the pioneering work on spherical DCAA, \cite{jiang20263d}  mainly focuses on the architecture design, beampattern characterization, and angular resolution analysis.
However,  how to fully exploit the unique characteristics of spherical DCAA for low-altitude ISAC remains largely unexplored.

In this paper, we further investigate the full 3D coverage of spherical DCAA for low-altitude ISAC systems.
Specifically, the sensing performance of spherical DCAA is first analyzed, where the close-form expressions of sensing signal-to-noise ratio (SNR), area average probability of detection, and the Cramer-Rao lower bounds (CRLBs) for both elevation and azimuth angle estimation are derived.
Furthermore, a low-altitude ISAC optimization problem is formulated to maximize the worst-case sensing SNR over a prescribed aerial sensing region while guaranteeing the communication quality-of-service (QoS) requirements of ground users.
To efficiently solve the resulting mixed-integer non-convex problem, a novel greedy-based joint array selection and beamforming optimization framework is developed.

The main contributions of this paper are summarized as follows.

\begin{itemize}

\item First, the 3D sensing performance of spherical DCAA is characterized, where the conventional UPA employed with Kronecker product codebook (KPC) serves as the benchmark.
    Specifically, a closed-form expression of the maximum achievable sensing SNR is derived, which reveals that the spherical DCAA can achieve more uniform and higher sensing SNR distribution over the prescribed region.
    To further quantify the sensing coverage advantage of spherical DCAA, the area average probability of detection is analyzed.
    Moreover, the approximate closed-form expressions of the CRLBs for both azimuth and elevation angle estimation are derived, which reveals that spherical DCAA enjoys substantially lower CRLBs over the entire angular region than the conventional UPA.

\item Second, a low-altitude ISAC optimization problem is formulated to maximize the worst-case sensing SNR within a prescribed aerial sensing region while satisfying the communication signal-to-interference-plus-noise ratio (SINR) requirements of ground users.
    To solve the resulting mixed-integer non-convex problem, a greedy-based joint array selection and beamforming framework is developed.
    Furthermore, a closed-form structural solution is first derived for a special orthogonal-channel case, providing useful design insights and enabling a low-complexity implementation for general multi-user communication and regional sensing scenarios.

\item Third, extensive simulation results are provided to validate the effectiveness of spherical DCAA for low-altitude ISAC.
    The results show that spherical DCAA achieves a superior communication--sensing SNR tradeoff than conventional UPA, thanks to the enhanced energy-focusing capability enabled by directional antenna elements.
    Furthermore, owing to its more uniform and superior angular resolution, spherical DCAA provides significantly improved aerial sensing coverage over the prescribed sensing region.

\end{itemize}

The rest of this paper is organized as follows.
Section~\ref{system model} first breifly introduce spherical DCAA and describe the considered low-altitude ISAC system.
Section~\ref{sensing performance analysis} analyzes the sensing performance of spherical DCAA.
In Section~\ref{optimization problem}, a joint array selection and beamforming optimization problem is considered for low-altitude ISAC, where a novel greedy-based algorithm is proposed to solve such a non-convex optimization problem.
Simulation results and discussion are detailed in Section~\ref{simulation results}.
Finally, Section~\ref{conclusion} concludes this paper.

\section{System Model}\label{system model}
\subsection{Spherical DCAA Architecture}
\begin{figure}
    \centering
        \subfigure[]{
    \includegraphics[width=0.35\textwidth]{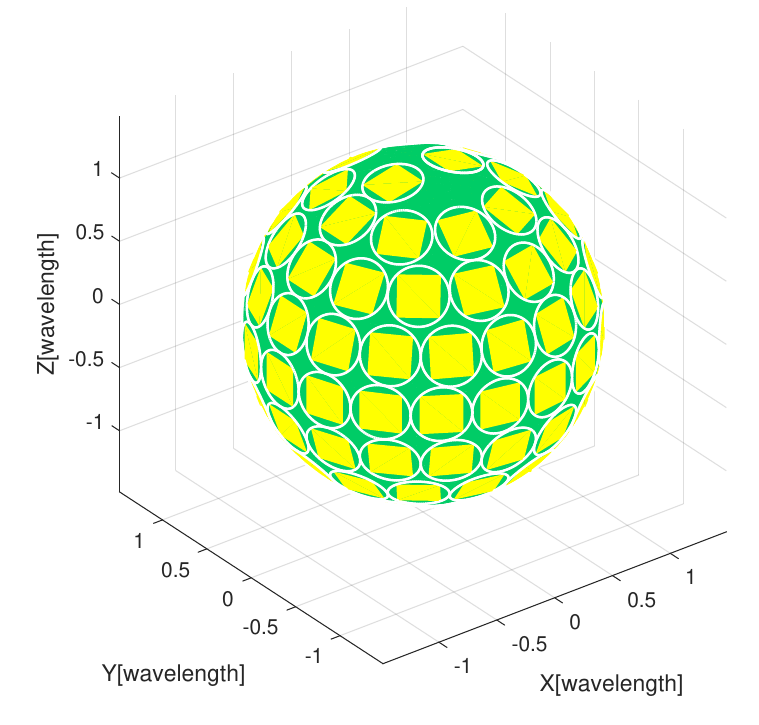}\label{SAA architecture}
    }
    \hspace{-15pt}
    \subfigure[]{
    \includegraphics[width=0.35\textwidth]{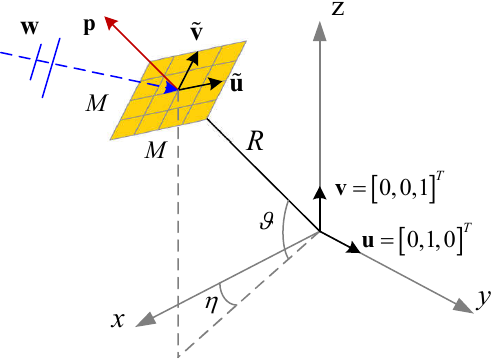}\label{sUPA}
    }
  \caption{An illustration of the spherical DCAA architecture and the placement of each sUPA in the corresponding 3D coordinate system.}\label{SAA}\vspace{-15pt}
\end{figure}
Fig.~\ref{SAA} gives a schematic illustration of the spherical DCAA architecture in 3D coordinate system.
The spherical DCAA consists of $N$ sUPAs arrange on the surface of a sphere.
Without loss of generality, we assume that each sUPA employs $M\times M$ antenna elements with half-wavelength spacing.
Different from the conventional UPA, all the antenna elements of each sUPA are directly connected to an analog combiner or splitter, without requiring any phase shifter.
Note that the sUPA has the naturally energy-focusing capability, even without relying on analog or digital beamforming.

As illustrated in Fig.~\ref{sUPA}, consider the incoming signal with the wave vector $\mathbf{w}=[-\cos\theta\cos\phi,-\cos\theta\sin\phi,-\sin\theta]^T$, where $\phi$ and $\theta$ denote the azimuth and elevation angles, respectively, with $-\pi\le \phi\le \pi$ and $-\pi/2\le \theta\le \pi/2$.
Let $\mathbf{u}=[0,1,0]^T$ and $\mathbf{v}=[0,0,1]^T$ be the reference direction vectors along the $y$- and $z$-axis, respectively.
The orientations of the sUPA with the normal vector $\mathbf{p}(\eta,\vartheta)=[\cos\vartheta\cos\eta,\cos\vartheta\sin\eta,\sin\vartheta]^T$ can be represented by
\begin{equation}
\begin{aligned}
&\tilde{\mathbf{u}}=\mathbf{R}(\eta,\vartheta)\mathbf{u}=[-\sin\eta,\cos\eta,0]^T,\\
&\tilde{\mathbf{v}}=\mathbf{R}(\eta,\vartheta)\mathbf{v}=[-\cos\eta\sin\vartheta,-\sin\eta\sin\vartheta,\cos\vartheta]^T,
\end{aligned}
\end{equation}
where $\mathbf{R}(\eta,\vartheta)$ is the rotation matrix without considering the self-spinning of the sUPA, $\eta\in[-\eta_{\max},\eta_{\max}]$ and $\vartheta\in[-\vartheta_{\max},\vartheta_{\max}]$ represent the yaw and pitch angles, rotating round the $z$- and $y$-axis, respectively, with $0\le \eta_{\max}\le \pi$ and $0\le \vartheta_{\max}\le \pi/2$.
The rotation matrix $\mathbf{R}(\eta,\vartheta)$ is defined as
\begin{equation}
\mathbf{R}(\eta,\vartheta)=\left[
\begin{matrix}
&\cos\eta & -\sin\eta  &0\\
&\sin\eta & \cos\eta  &0\\
&0  &0  &1
\end{matrix}
\right]
\left[
\begin{matrix}
&\cos\vartheta & 0 & -\sin\vartheta\\
&0 & 1 & 0\\
&\sin\vartheta & 0 & \cos\vartheta
\end{matrix}
\right].
\end{equation}

For the incoming signal with the wave vector $\mathbf{w}$, the array response vector of the rotated sUPA can be expressed as
\begin{equation}\label{array response0}
\bar{\mathbf{a}}(\tilde{\mathbf{u}},\tilde{\mathbf{v}},\mathbf{w})
=e^{j\omega_0}\sqrt{G(\tilde{\mathbf{u}},\tilde{\mathbf{v}},\mathbf{w})}
\mathbf{a}(\tilde{\mathbf{u}},\mathbf{w})\otimes\mathbf{a}(\tilde{\mathbf{v}},\mathbf{w}),
\end{equation}
where $\omega_0$ is a common phase shift from the reference antenna element, $G(\tilde{\mathbf{u}},\tilde{\mathbf{v}},\mathbf{w})$ accounts the radiation pattern of each antenna element of the sUPA, which can be also expressed as $G(\eta-\phi,\vartheta-\theta)$, and $\mathbf{a}(\tilde{\mathbf{u}},\mathbf{w})$ and $\mathbf{a}(\tilde{\mathbf{v}},\mathbf{w})$ are the normalized horizontal and vertical steering vectors of the rotated sUPA, respectively, which is given by $\mathbf{a}(\mathbf{x},\mathbf{y})=[1,e^{j\pi\mathbf{x}^T\mathbf{y}},\cdots,e^{j(M-1)\pi\mathbf{x}^T\mathbf{y}}]^T$.
Note that all $M\times M$ antenna elements of the sUPA are directly connected, thus the normalized beam pattern of the rotated sUPA is
\begin{equation}\label{array response}
\begin{aligned}
&r(\tilde{\mathbf{u}},\tilde{\mathbf{v}}, \mathbf{w})\\
&=\sqrt{1/M^2}\mathbf{1}^T_{M^2\times1}\bar{\mathbf{a}}(\tilde{\mathbf{u}},\tilde{\mathbf{v}},\mathbf{w})\\
&=e^{j\omega_0}\sqrt{M^2G(\tilde{\mathbf{u}},\tilde{\mathbf{v}},\mathbf{w})}\mathcal{H}_M(-\tilde{\mathbf{u}}^T\mathbf{w})\mathcal{H}_M(-\tilde{\mathbf{v}}^T\mathbf{w})\\
&\triangleq e^{j\omega_0}\sqrt{M^2G(\eta-\phi,\vartheta-\theta)}f(\phi,\theta;\eta,\vartheta),
\end{aligned}
\end{equation}
where $\mathcal{H}_M(x)\triangleq e^{j{\pi}/{2}(M-1)x}\frac{\sin({\pi}/{2}Mx)}{M\sin({\pi}/{2}x)}$ is the Dirichlet kernel function and $f(\phi,\theta;\eta,\vartheta)$ is defined as
\begin{equation}
\begin{aligned}
f(\phi,\theta;\eta,\vartheta)\triangleq&\mathcal{H}_M(\cos\theta\sin(\phi-\eta))\\
&\times\mathcal{H}_M(\sin\theta\cos\vartheta-\cos\theta\sin\vartheta\cos(\phi-\eta)).
\end{aligned}
\end{equation}
To facilitate the theoretical analysis, we make the following reasonable assumptions for the antenna pattern $G(\phi,\theta)$:
\begin{itemize}
  \item The 3dB beamwidth of $G(\phi,\theta)$, denoted by $(\phi_{\text{3dB}},\theta_{\text{3dB}})$, is wider than the angular resolution defined by the sUPA spacing, for ensuring that $G(\phi,\theta)$ varies slowly compared to the array factor $f(\phi,\theta;\eta,\vartheta)$ within the mainlobe.
  \item $G(\phi,\theta)$ is symmetric with respect to (w.r.t.) the boresight and achieves its global maximum at $(\phi,\theta)=(0,0)$.
  \item $G(\phi,\theta)$ is monotonically non-increasing w.r.t. $(\phi,\theta)$ within the mainlobe for $|\phi|\le \phi_{\text{3dB}}/2$ and $|\theta|\le \theta_{\text{3dB}}/2$.
\end{itemize}
Note that the transmit response vector for steering a signal towards $(\phi,\theta)$ can be similarly derived based on \eqref{array response}.
Thus, the details are omitted for brevity.
Under the above assumptions, we have the following remarks.

{\it Remark 1}:
The beampattern of sUPA in \eqref{array response} demonstrates the energy-focusing capability even without relying on any analog or digital beamforming.
For any incoming/outgoing signal from/to a particular direction $(\phi,\theta)$, only the sUPA rotated aligned with $(\phi,\theta)$ could receive/steering significant power, while the power of all other sUPAs whose boresights deviate from $(\phi,\theta)$ is almost negligible.

%Therefore, a flexible beamforming can be achieved by selecting sUPAs with appropriate rotations to connect with the RF chains for further baseband digital processing.
%% as illustrated in Fig.~\ref{SAA system}.
%Note that different from traditional HBF architectures that typically rely on a large number of phase shifters, the spherical DCAA architecture employs an array selection network (ASN) composed of low-cost RF switches, which may enjoys the hardware cost efficient.
%Denote by $N_{\text{RF}}$ the number of RF chains, with $N_{\text{RF}}< N$.
%To dynamically select $N_{\text{RF}}$ out of $N$ sUPAs to connect with the RF chains, the array selection matrix is defined as $\mathbf{U}\in\mathbb{R}^{N_{\text{RF}}\times N}$, which is binary matrix, satisfying that $\|[\mathbf{U}]_{i,:}\|_0=1$ and $\|[\mathbf{U}]_{:,j}\|_0\le 1$ for $i=1,\cdots,N_{\text{RF}}$ and $j=1,\cdots,N$.
% \begin{figure}
%     \centering
%     {
%     \includegraphics[width=0.45\textwidth]{SAA system.png}\label{SAA architecture}
%     }
%   \caption{Spherical DCAA architecture without any phase shifters.}\label{SAA system}\vspace{-15pt}
% \end{figure}

{\it Remark 2}:
The azimuth and elevation angular resolutions, defined as the half of the null-to-null beamwidth of the mainlobe in \eqref{array response}, are respectively given by
\begin{equation}
\Delta\phi(\vartheta) = \arcsin\left(\frac{2}{M\cos\vartheta}\right), \
\Delta\theta = \arcsin\left(\frac{2}{M}\right),
\end{equation}
where the elevation angular resolution is independent from the incoming signal direction $(\theta,\phi)$, while although the azimuth angular resolution increases with the elevation angle $\theta$, it has the uniform azimuth angular resolution when the elevation angle is fixed, which is different from the conventional UPA.
It implies that the spherical DCAA achieves a more uniform and superior resolution than conventional UPA, especially at the high-elevation regions \cite{jiang20263d}.

%To effectively suppress the interference between adjacent sUPAs, the orientations of all $N$ sUPAs are designed based on the elevation and azimuth angular resolutions, i.e.,
%\begin{equation}
%\begin{aligned}
%&\vartheta_q = q\arcsin\left(\frac{2}{M}\right), q=0,\pm 1,\cdots, \pm N_e,\\
%&\eta_{p,q} = p\arcsin\left(\frac{2}{M\cos\vartheta_q}\right), p=0, \pm 1, \cdots, \pm N_a(\vartheta_q),
%\end{aligned}
%\end{equation}
%where $N_e=\left\lfloor{\theta_{\max}}/{\arcsin\left(\frac{2}{M}\right)}\right\rfloor$ and $N_a(\vartheta_q)=\left\lfloor{\phi_{\max}}/{\arcsin\left(\frac{2}{M\cos\vartheta_q}\right)}\right\rfloor
%$, with $\theta_{\max}$ and $\phi_{\max}$ being the maximum elevation and azimuth angles of the signal, respectively, and it satisfies that $\sum\nolimits_{q=-N_e}^{N_e}N_a(\vartheta_q)=N$.
%For ease of exposition, the two-dimensional orientation index $(\eta_{p,q},\vartheta_q)$ is transformed into one-dimensional version as $(\eta_n,\vartheta_n)$, $n\in\mathcal{N}\triangleq\{1,\cdots,N\}$, satisfying that $n=p+N_a(\vartheta_q)+1+\sum\nolimits_{k=-N_e}^{q-1}(2N_a(\vartheta_k)+1)$.
%Moreover, to avoid the signal blockage among sUPAs, the minimum radius $R$ of the sphere is $R_{\min}=\frac{M\lambda}{2\sqrt{2}\tan(0.5\arcsin(2/M))}$, with $\lambda$ being the signal wavelength \cite{jiang20263d}.
%Therefore, the spherical DCAA architecture can be parameterized by $\{M^2,N,R,(\eta_n,\vartheta_n), n\in\mathcal{N}\}$.

{\it Remark 3}:
According to \eqref{array response}, the 3D angular coverage is naturally divided by the spherical DCAA, where
each sUPA is only responsible for a small portion of the total coverage.
Thus, spherical DCAA offers a new design degree-of-freedom (DoF) to configure the radiation pattern $G_n(\cdot)$ of each sUPA~$n$ to meet different performance requirements.

The aforementioned advantages render the spherical DCAA particular appealing for the emerging low-altitude ISAC applications.
In such scenarios, communication users or aerial targets easily traverse the coverage hole of the BS equipped with traditional planar arrays, leading to the degraded communication and sensing performance.
In contrast, the spherical DCAA enables full 3D coverage and superior and uniform angular resolution.
This thus motivate our current work.

\begin{figure}
    \centering
    {
    \includegraphics[width=0.45\textwidth]{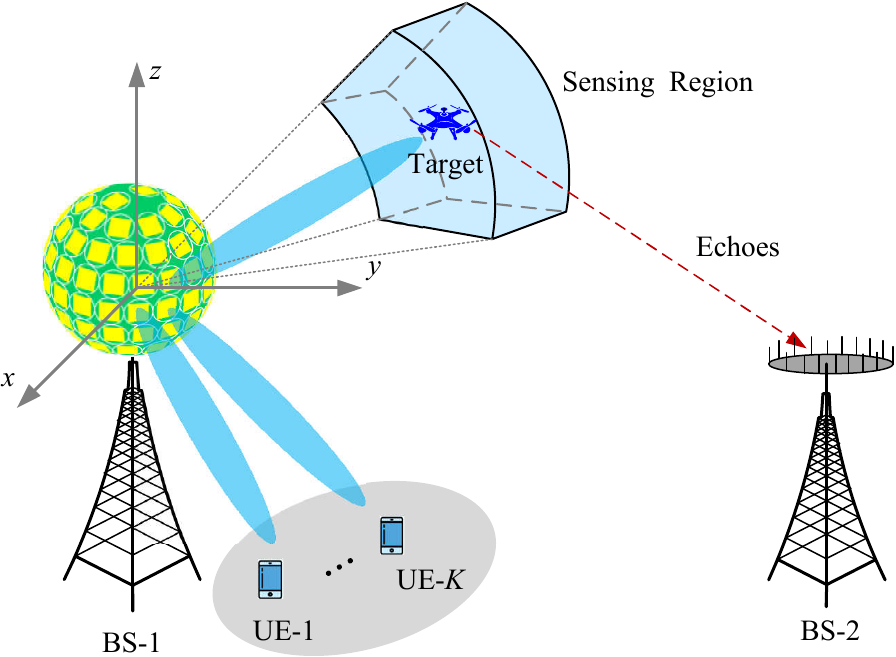}
  \caption{Low-altitude ISAC system based on spherical DCAA.}\label{SAA ISAC system}
    }\vspace{-15pt}
\end{figure}
\subsection{Spherical DCAA-based Low-Altitude ISAC}
As shown in Fig.~\ref{SAA ISAC system}, we consider a spherical DCAA-based low-altitude ISAC system, where BS-1 equipped with the spherical DCAA wishes to communicate with $K$ ground single-antenna user equipments (UEs) while simultaneously providing sensing coverage to the prescribed region $\mathcal{A}\in\mathbb{R}^3$, by cooperating with the BS-2\footnote{As the transmit signals of BS-1 can be known by BS-2 via backhual, they can serve as the bi-static radar for target sensing.}.
Consider the system operated at the mmWave frequency band and under the far-field wireless propagation assumption.

\subsubsection{Communication Model}
Let $\mathcal{K}=\{1,\cdots,K\}$ be the index set of all $K$ UEs.
The multi-path downlink communication channel from the $n$th sUPA to the $k$th UE is given by
\begin{equation}
\mathbf{h}_{nk} = \sum\nolimits_{l=1}^{L_k}\beta_{kl}\bar{\mathbf{a}}(\mathbf{u}_n,\mathbf{v}_n,\mathbf{w}_{kl}), n\in\mathcal{N}, k\in\mathcal{K},
\end{equation}
where $L_k$ is the number of multi-paths for user $k$, $\beta_{kl}$ denotes the complex-valued path gain for the $l$th path, $\mathbf{w}_{kl}$ is the normalized wave vector of the $l$th path, which is given by $\mathbf{w}_{kl}=[\cos\theta_{kl}\cos\phi_{kl},\cos\theta_{kl}\sin\phi_{kl},\sin\theta_{kl}]^T$, with $\phi_{kl}$ and $\theta_{kl}$ being the elevation and azimuth angle-of-departure (AoD), respectively, and $\bar{\mathbf{a}}(\mathbf{u}_n,\mathbf{v}_n,\mathbf{w}_{kl})$ is the array response vector of the $n$th sUPA as given by \eqref{array response0}, with the orientation vectors $\mathbf{u}_n=\mathbf{R}(\eta_n,\vartheta_n)\mathbf{u}$ and $\mathbf{v}_n=\mathbf{R}(\eta_n,\vartheta_n)\mathbf{v}$.
Note that to avoid the blockage, the transmit paths must lie on the same side of the sUPA.
This yields the following constraint, $\mathbf{w}_{kl}^T\mathbf{p}(\eta_n,\vartheta_n) > 0$, $l=1,\cdots,L_k$.
Therefore, for the spherical DCAA, the downlink communication channels from all $N$ sUPAs to the $k$th UE can be compactly expressed as
\begin{equation}\label{downlink channel}
\bar{\mathbf{h}}_k = [\mathbf{h}_{1k}^T,\cdots,\mathbf{h}_{Nk}^T]^T = \sum\nolimits_{l=1}^L\beta_{kl}\mathrm{vec}(\mathbf{A}(\mathbf{w}_{kl})),
\end{equation}
where $\bar{\mathbf{h}}_k\in\mathbb{C}^{M^2N\times1}$ and  $\mathbf{A}(\mathbf{w}_{kl})\triangleq [\bar{\mathbf{a}}(\mathbf{u}_n,\mathbf{v}_n,\mathbf{w}_{kl})]_{n\in\mathcal{N}}$.
Here, we assume that all $K$ UEs have the same number of multi-paths, i.e., $L_k=L$, $\forall k$, for notation simplicity.

%Therefore, a flexible beamforming can be achieved by selecting sUPAs with appropriate rotations to connect with the RF chains for further baseband digital processing.
%%% as illustrated in Fig.~\ref{SAA system}.
%%Note that different from traditional HBF architectures that typically rely on a large number of phase shifters, the spherical DCAA architecture employs an array selection network (ASN) composed of low-cost RF switches, which may enjoys the hardware cost efficient.
%Denote by $N_{\text{RF}}$ the number of RF chains, with $N_{\text{RF}}< N$.
%To dynamically select $N_{\text{RF}}$ out of $N$ sUPAs to connect with the RF chains, the array selection matrix is defined as $\mathbf{U}\in\mathbb{R}^{N_{\text{RF}}\times N}$, which is binary matrix, satisfying that $\|[\mathbf{U}]_{i,:}\|_0=1$ and $\|[\mathbf{U}]_{:,j}\|_0\le 1$ for $i=1,\cdots,N_{\text{RF}}$ and $j=1,\cdots,N$.

Denote by $\mathbf{s}=[s_1,\cdots,s_K]^T$ the transmit information-bearing symbol vector for all $K$ UEs, with $\mathbb{E}[\mathbf{s}\mathbf{s}^H]=\mathbf{I}_K$.
The digital transmit beamforming matrix of the BS-1 is given by $\mathbf{F}=[\mathbf{f}_1,\cdots,\mathbf{f}_K]\in\mathbb{C}^{N_{\text{RF}}\times K}$, where $\mathbf{f}_k\in\mathbb{C}^{N_{\text{RF}}\times1}$ denotes the transmit beamforming vector for the $k$th UE, with $N_{\rm RF}$ being the number RF chains.
Here, it considers that $K\le N_{\text{RF}}<N$.
The baseband processing output is $\mathbf{x}=\mathbf{F}\mathbf{s}$.
After that, the array selection network (ASN) is introduced to dynamically select $N_{\text{RF}}$ out of $N$ sUPAs to connect with the RF chains for transmission.
The array selection matrix is defined as $\mathbf{U}\in\mathbb{R}^{N_{\text{RF}}\times N}$, which is binary matrix, satisfying that $\|[\mathbf{U}]_{i,:}\|_0=1$ and $\|[\mathbf{U}]_{:,j}\|_0\le 1$ for $i=1,\cdots,N_{\text{RF}}$ and $j=1,\cdots,N$.
The transmit output of the ASN is
\begin{equation}
\bar{\mathbf{x}}=\mathbf{U}^H\mathbf{x}=\mathbf{U}^H\mathbf{F}\mathbf{s}.
\end{equation}

For signal transmission, the power splitter is directly connected with all $M\times M$ antenna elements of each sUPA.
Finally, the equivalent transmit signal of the spherical DCAA for  all $M^2N$ antenna elements is
\begin{equation}
\mathbf{x}_t = \bar{\mathbf{x}}\otimes\sqrt{1/M^2}\mathbf{1}_{M^2\times1}.
\end{equation}
The transmit power of $\mathbf{x}_t$ satisfies that
\begin{equation}
\begin{aligned}
\mathbb{E}[\|\mathbf{x}_t\|^2]&=\mathbb{E}\left[\left\|\bar{\mathbf{x}}\otimes\sqrt{1/M^2}\mathbf{1}_{M^2\times1}\right\|^2\right]\\
&=\mathbb{E}[\|\bar{\mathbf{x}}\|^2]\cdot\left\|\sqrt{1/M^2}\mathbf{1}_{M^2\times1}\right\|^2\\
&=\mathbb{E}[\|\mathbf{U}^H\mathbf{F}\mathbf{s}\|^2]\overset{(a)}{=}\sum\nolimits_{k=1}^{K}\|\mathbf{f}_k\|^2= P_t,
\end{aligned}
\end{equation}
where $P_t$ denotes the total transmit power and $(a)$ holds since $\mathbf{U}\mathbf{U}^H=\mathbf{I}_{N_{\text{RF}}}$ and $\mathbf{E}[\mathbf{s}\mathbf{s}^H]=\mathbf{I}_K$.

With the downlink communication channel in \eqref{downlink channel}, the received signal at the $k$th UE is
\begin{equation}
\begin{aligned}
  y_k &= \bar{\mathbf{h}}_k^H\mathbf{x}_t + z_k\\
      &=\big(\sum\nolimits_{l=1}^L\beta_{kl}\mathrm{vec}(\mathbf{A}(\mathbf{w}_{kl}))\big)^H\big(\bar{\mathbf{x}}\otimes\sqrt{{1}/{M^2}}\mathbf{1}_{M^2\times1}\big) + z_k\\
     &=\big(\sum\nolimits_{l=1}^{L}\beta_{kl}\sqrt{1/M^2}\mathbf{1}_{M^2\times1}\mathbf{A}(\mathbf{w}_{kl})\big)^H\bar{\mathbf{x}} + z_k\\
     &=
     \big(\sum\nolimits_{l=1}^{L}\beta_{kl}\mathbf{r}(\mathbf{w}_{kl})\big)^H
     \bar{\mathbf{x}} + z_k,
\end{aligned}
\end{equation}
where $\mathbf{r}(\mathbf{w}_{kl})\triangleq[r(\mathbf{u}_1,\mathbf{v}_1,\mathbf{w}_{kl}),\cdots,r(\mathbf{u}_N,\mathbf{v}_N,\mathbf{w}_{kl})]^T\in\mathbb{C}^{N\times1}$ with $r(\mathbf{u}_n,\mathbf{v}_n,\mathbf{w}_{kl})$, $n\in\mathcal{N}$ according to \eqref{array response}, and $z_k\sim\mathcal{CN}(0,\sigma^2)$ is the additive white Gaussian noise (AWGN).
Denote by $\bar{\mathbf{h}}_{ek}\triangleq\sum\nolimits_{l=1}^{L}\beta_{kl}\mathbf{r}(\mathbf{w}_{kl})$ the equivalent channel from $N$ sUPAs to the $k$th UE.
The received SINR of the $k$th UE is
\begin{equation}\label{communication SINR}
\gamma_k(\mathbf{U},\mathbf{F}) = \frac{\left|\bar{\mathbf{h}}_{ek}^H\mathbf{U}^H\mathbf{f}_k\right|^2}{\sum\limits_{i\neq k}\left|\bar{\mathbf{h}}_{ek}^H\mathbf{U}^H\mathbf{f}_i\right|^2+\sigma^2}, k\in\mathcal{K}.
\end{equation}

\subsubsection{Sensing Model}
The sensing coverage region is defined as $\mathcal{A}=\mathcal{D}\times{\Phi}\times{\Theta}$, with $\mathcal{D}=[d_{\min},d_{\max}]$, $\Phi=[\phi_{\min},\phi_{\max}]$, and $\Theta=[\theta_{\min},\theta_{\max}]$.
Consider a point-like target within the sensing coverage region $\mathcal{A}$, whose location can be expressed  $\mathbf{t}=d_s\mathbf{w_s}$, where $\mathbf{w}_s\triangleq[\cos\theta_s\cos\phi_s,\cos\theta_s\sin\phi_s,\sin\theta_s]^T$ denotes the normalized wave vector towards $(\phi_s,\theta_s)$ and $d_s$ is the rational distance from the target to BS-1, with $d_s\in\mathcal{D}$, $\phi_s\in\Phi$, and $\theta_s\in\Theta$, respectively.

For bistatic sensing, denote by $(\phi_s',\theta_s')$ the AoA from the target to BS-2.
The echo signal received at BS-2 is
\begin{equation}
\begin{aligned}
\mathbf{y}_s(\mathbf{t}) &= \alpha_s(\mathbf{t})\mathbf{b}(\phi_s',\theta_s')\mathbf{r}(\mathbf{w}_s)^H\bar{\mathbf{x}} + \mathbf{z}_s\\
&=\alpha_s(\mathbf{t})\mathbf{b}(\phi_s',\theta_s')\mathbf{r}(\mathbf{w}_s)^H\mathbf{U}^H\mathbf{F}\mathbf{s}+\mathbf{z}_s,
\mathbf{t}\in\mathcal{A},
\end{aligned}
\end{equation}
where $\alpha_s(\mathbf{t})\triangleq\sqrt{\frac{\delta^2_{\text{RCS}}\lambda^2}{4\pi(4\pi\|\mathbf{t}\|\cdot\|\mathbf{t}-\mathbf{o}_2\|)^2}}e^{j\omega_s}$
is the complex channel coefficient, with $\delta_{\text{RCS}}$, $\mathbf{o}_2\in\mathbb{R}^3$, and $\omega_s$ being the radar cross section (RCS) of the target, the location of BS-2, and the phase rotation, respectively, $\mathbf{b}(\phi_s',\theta_s')\in\mathbb{C}^{N_R\times1}$ is the normalized receive steering vector, and $\mathbf{z}_s$ is the AWGN vector, satisfying that $\mathbb{E}[\mathbf{z}_s\mathbf{z}_s^H]=\sigma^2_s\mathbf{I}_{N_R}$, where $N_R$ is the number of receive antenna of the BS-2.

Note that since the information-bearing symbols $\mathbf{s}$ can be prior known by the cooperative BS-2, the randomness of the received echoes can be removed.
Moreover, to highlight the potentials of spherical DCAA for sensing, we focus on the AoD $(\phi_s,\theta_s)$ estimation and assume that an optimal receiver beamformer, denoted by $\mathbf{g}=\mathbf{b}(\theta_s',\phi_s')$, is applied at BS-2.
Thus, the resulting received signal is
\begin{equation}\label{sensing rx}
y_s(\mathbf{t}) =\sum\nolimits_{k=1}^{K} \alpha_s(\mathbf{t})\mathbf{r}(\mathbf{w}_s)^H\mathbf{U}^H\mathbf{f}_k + z_s, \mathbf{t}\in\mathcal{A}.
\end{equation}
The received sensing SNR for a target located at $\mathbf{t}\in\mathcal{A}$ is
\begin{equation}\label{sensing_SNR}
\gamma_s(\mathbf{U},\mathbf{F};\mathbf{t}) =\sum\nolimits_{k=1}^{K}\left|\alpha_s(\mathbf{t})\mathbf{r}(\mathbf{w}_s)^H\mathbf{U}^H\mathbf{f}_k\right|^2/\sigma_s^2.
\end{equation}

From \eqref{communication SINR} and \eqref{sensing_SNR}, it observes that both communication and sensing performance critically depend on the array selection and digital beamforming matrix, i.e., $\mathbf{U}$ and $\mathbf{F}$.
In the following, we first analyze the spherical DCAA-based sensing performance.
After that, a joint array selection and beamforming optimization problem is considered to simultaneously guaranteeing both communication and sensing performance.

\section{Sensing Performance Analysis}\label{sensing performance analysis}
In this section, we first analyze the sensing performance of the spherical DCAA.
For fairness, we consider the UPA composed of $M\times M$ antenna elements as the benchmark, whose normal vector is aligned with the positive $y$-axis.

The steering vector of the UPA towards the target direction $(\phi_s,\theta_s)$ can be expressed as
\begin{equation}\label{UPA response}
\mathbf{a}_{\text{UPA}}(\phi_s,\theta_s) = \mathbf{a}(\theta_s)\otimes\mathbf{a}(\theta_s,\phi_s),
\end{equation}
where $\mathbf{a}(\theta_s)=[1,e^{j\pi\sin\theta,\cdots,e^{j\pi(M-1)\sin\theta_s}}]^T$ and $\mathbf{a}(\theta_s,\phi_s)=[1,e^{j\pi\cos\theta_s\sin\phi_s},\cdots,e^{j\pi(M-1)\cos\theta_s\sin\phi_s}]^T$.
For ease of implementation, the conventional KPC is considered for transmit beamforming.
The $(p,q)$-th codeword of the KPC is given by $\mathbf{f}_{p,q}=\mathbf{f}_q\otimes\mathbf{f}_p$, with
\begin{equation}\label{KPC}
\begin{aligned}
&\mathbf{f}_q={1}/{\sqrt{M}}[1,e^{j\pi\frac{2q}{M}},\cdots,e^{j\pi(M-1)\frac{2q}{M}}]^T,\\
&\mathbf{f}_p={1}/{\sqrt{M}}[1,e^{j\pi\frac{2p}{M}},\cdots,e^{j\pi(M-1)\frac{2p}{M}}]^T,
\end{aligned}
\end{equation}
where $q\in\{-\frac{N_v-1}{2},\cdots,\frac{N_v-1}{2}\}$, $p\in\{-\frac{N_h-1}{2},\cdots,\frac{N_h-1}{2}\}$, with $N_v=2\lfloor\frac{M\sin\theta_{\max}}{2}\rfloor+1$ and $N_h=2\lfloor\frac{M\sin\phi_{\max}}{2}\rfloor+1$ being the number of elevation and azimuth codewords, respectively.
%The received signal is
%\begin{equation}
%y_{s,\rm UPA}(\mathbf{t}) = \alpha_s(\mathbf{t})\mathbf{a}^H_{\rm UPA}(\phi_s,\theta_s)\mathbf{f}_{p,q} + z_s, \mathbf{t}\in\mathcal{A}.
%\end{equation}

\subsection{Sensing SNR}\label{sensing SNR analysis}
\subsubsection{Spherical DCAA}
To maximize the sensing SNR of the spherical DCAA for a target located at $\mathbf{t}\in\mathcal{A}$, disregarding the communication constraints, a single RF chain is sufficient.
For such scenarios, the problem becomes to select the optimal sUPA $n^*\in\mathcal{N}$, whose normal vector $\mathbf{p}(\eta_{n^*},\vartheta_{n^*})$ aligns most closely with the wave vector $\mathbf{w}_s$, with $n^*=\arg\max_{n\in\mathcal{N}}(\mathbf{w}_s^T\mathbf{p}(\eta_n,\vartheta_n))$.
Therefore, according to \eqref{array response} and \eqref{sensing_SNR}, the maximum achievable sensing SNR is
\begin{equation}\label{maximum sensing_SNR}
\begin{aligned}
\gamma_{s}(\mathbf{t}) =& |\alpha_s(\mathbf{t})|^2\bar{P}_tM^2G_{n^*}(\eta_{n^*}-\phi_s,\vartheta_{n^*}-\theta_s)\\
&\times|f(\phi_s,\theta_s;\eta_{n^*},\vartheta_{n^*})|^2,
\end{aligned}
\end{equation}
where $\bar{P}_t\triangleq P_t/\sigma_s^2$.
Note that the sensing SNR depends on the target location $\mathbf{t}$, parameterized by $\{d_s,\phi_s,\theta_s\}$.
Since the path-loss factor $|\alpha_s(\mathbf{t})|^2$ is governed by the product of the bistatic ranges, i.e., $\|\mathbf{t}\|\cdot\|\mathbf{t}-\mathbf{o}_2\|$, to characterize the angular sensitivity of the spherical DCAA independent of $|\alpha_s(\mathbf{t})|^2$,
we analyze the sensing SNR on a specific Cassini oval $\mathcal{A}_c\subseteq\mathcal{A}$~\cite{kuschel2019tutorial}.
For $\mathbf{t}\in\mathcal{A}_c$, it has $\|\mathbf{t}\|\cdot\|\mathbf{t}-\mathbf{o}_2\|=C_0$, with $C_0$ being a constant, and thus $\alpha_s(\mathbf{t})=\alpha_s$.

In this case, the sensing SNR primarily depends on the AoD $(\phi_s,\theta_s)$ associated with the target location $\mathbf{t}$.
When $(\phi_s,\theta_s)$ is aligned with the boresight of the $n^*$th sUPA, such that $\phi_{s}-\eta_{n^*}=0$ and $\theta_s-\vartheta_{n^*}=0$, \eqref{maximum sensing_SNR} reduces to
\begin{equation}\label{best sensing SNR}
\gamma_{s,\text{best}} = |\alpha_s|^2\bar{P}_tM^2G_{n^*}(0,0),
\end{equation}
which corresponds to the best case.
In contrast, the worst case occurs when the target is located at the edge of the mainlobe, i.e., $|\phi_s-\eta_{n^*}|=\Delta\phi(\vartheta_n^*)/2$ and $|\theta_s-\vartheta_{n^*}|=\Delta\theta/2$.
The maximum achievable sensing SNR for the worst case is
\begin{equation}\label{worst sensing SNR}
\small
\begin{aligned}
\gamma_{s,\text{worst}} =& |\alpha_s|^2\bar{P}_tM^2G_{n^*}\left(\frac{\Delta\phi(\vartheta_{n^*})}{2},\frac{\Delta\theta}{2}\right)\\
&\times|f(\phi_s,\theta_s;\eta_{n^*},\vartheta_{n^*})|^2.
\end{aligned}
\end{equation}
To ensure seamless sensing coverage, the sensing SNR for the worst case should be no smaller than $\gamma_{th}$, i.e., $\gamma_{s,\text{worst}}\ge \gamma_{th}$.
Thus, the antenna element pattern $G_{n^*}(\cdot,\cdot)$ should satisfy that
\begin{equation}
G_{n^*}(\Delta\phi(\vartheta_{n^*})/2,\Delta\theta/2) \ge \frac{\gamma_{th}}{\xi|f(\phi_s,\theta_s;\eta_{n^*},\vartheta_{n^*})|^2},
\end{equation}
where $\xi\triangleq|\alpha_s|^2\bar{P}_tM^2$, and the 3dB beamwidth of $G_{n^*}(\cdot,\cdot)$ should be no smaller than the angular resolution, i.e., $\phi_{\text{3dB}}\ge \Delta\phi(\vartheta_{n^*})$ and $\theta_{\text{3dB}}\ge \Delta\theta$.

\subsubsection{UPA}
Similar to \eqref{maximum sensing_SNR}, when the codeword $\mathbf{f}_{p^*,q^*}$ of the KPC aligned with the steering vector $\mathbf{a}_{\text{UPA}}(\phi_s,\theta_s)$ of UPA, the resulting sensing SNR incorporating the antenna pattern is
\begin{equation}\label{maximum sensing_SNR UPA}
\begin{aligned}
&\gamma_{\text{UPA}}(\phi_s,\theta_s;p^*,q^*)\\
&\quad=|\alpha_s|^2\bar{P}_tM^2G_{\text{UPA}}(\phi_s,\theta_s)\left|\mathbf{a}_{\text{UPA}}^H(\phi_s,\theta_s)\mathbf{f}_{p^*,q^*}\right|^2\\
&\quad=|\alpha_s|^2\bar{P}_tM^2G_{\text{UPA}}(\phi_s,\theta_s)\times\\
&\qquad\underbrace{
  |\mathcal{H}_M(\cos\theta_s\sin\phi_s-{2p^*}/{M})\mathcal{H}_M(\sin\theta_s-{2q^*}/{M})|^2
}_{|f_{\text{BF}}(\phi_s,\theta_s;p^*,q^*)|^2},
\end{aligned}
\end{equation}
where $G_{\text{UPA}}(\phi_s,\theta_s)$ denotes the antenna pattern of the UPA towards the AoD $(\phi_s,\theta_s)$, which only peaks at $(\phi_s,\theta_s)=(0,0)$, and $|f_{\text{BF}}(\phi_s,\theta_s;p^*,q^*)|^2$ represents the transmit beampattern.
Thus, for the best case, i.e., $\sin\theta_s=\frac{2q^*}{M}$ and $\sin\phi_s=\frac{2p^*}{M\cos\theta_s}=\frac{2p^*}{M\sqrt{1-\left(\frac{2q^*}{M}\right)^2}}$, \eqref{maximum sensing_SNR UPA} reduces to
\begin{equation}\label{best case UPA}
\gamma_{\text{UPA,best}}(\phi_s,\theta_s) = |\alpha_s|^2\bar{P}_tM^2G_{\text{UPA}}(\phi_s,\theta_s).
\end{equation}
On the other hand, similar to \eqref{worst sensing SNR}, when $|\sin\theta_s-\frac{2q^*}{M}|=\frac{1}{M}$ and $|\cos\theta_s\sin\theta_s-\frac{2p^*}{M}|=\frac{1}{M}$, the worst case of maximum achievable sensing SNR for UPA is derived as
\begin{equation}\label{worst sensing SNR UPA}
  \gamma_{\text{UPA,worst}}(\phi_s,\theta_s) = |\alpha_s|^2\bar{P}_tM^2G_{\text{UPA}}(\phi_s,\theta_s)|\mathcal{H}_M(1/M)|^4.
\end{equation}

Note that as the boresight of UPA is always aligned with the positive $y$-axis, to guarantee the seamless coverage, the 3dB beamwidth of UPA antenna pattern, denoted by $(\phi_{\text{3dB,UPA}},\theta_{\text{3dB,UPA}})$, should cover the whole angular region of interest, i.e., $\phi_{\text{3dB,UPA}}\ge \phi_{\max}-\phi_{\min}$ and $\theta_{\text{3dB,UPA}}\ge \theta_{\max}-\theta_{\min}$.
In contrast, for the spherical DCAA, since each sUPA only needs to cover a portion of the whole region, directional antenna elements with enhanced energy-focusing capability can be applied for improving the sensing SNR, while guaranteeing the seamless sensing coverage.

Specifically, given the sum of antenna gain as $G_{\text{sum}}=\int_{-\infty}^{\infty}\int_{-\infty}^{\infty} G(\phi,\theta)\mathrm{d}\phi\mathrm{d}\theta$, the peak antenna gain can be approximated expressed as $G(0,0)\approx G_{\text{sum}}/(\epsilon\phi_{\text{3dB}}\theta_{\text{3dB}})$ \cite{balanis2016antenna}, where $\epsilon$ is a positive factor.
As the required 3dB beamwidth of UPA is significantly wider than that of the spherical DCAA, with a fixed $G_{\text{sum}}$, it follows that
\begin{equation}
\frac{G_n(0,0)}{G_{\text{UPA}}(0,0)}=\frac{\phi_{\text{3dB,UPA}}\theta_{\text{3dB,UPA}}}{\phi_{\text{3dB},n}\theta_{\text{3dB},n}}\gg 1,
\end{equation}
which implies that the spherical DCAA can achieve higher beamforming gain than conventional UPA.

\begin{figure}
    \centering
    {
    \includegraphics[width=0.48\textwidth]{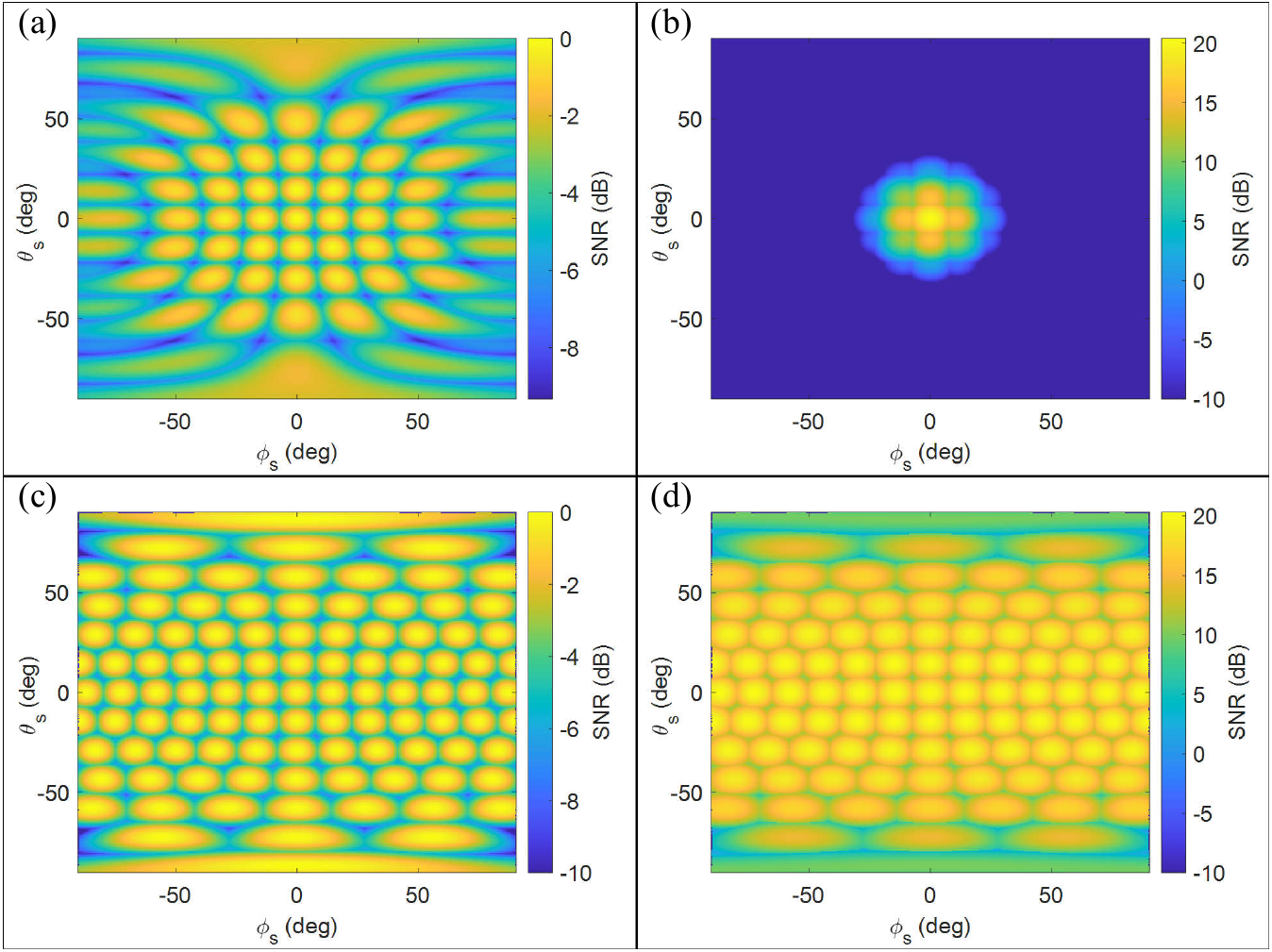}
    }
  \caption{Comparison of the sensing SNR distributions. (a) UPA with omnidirectional antenna. (b) UPA with directional antenna. (c) Spherical DCAA with omnidirectional antenna. (d) Spherical DCAA with directional antenna.}\label{Sensing SNR comparison}\vspace{-15pt}
\end{figure}
Fig.~\ref{Sensing SNR comparison} compares the sensing SNR distributions of the conventional UPA and the spherical DCAA with omnidirectional and directional antenna patterns, respectively.
It observes that the conventional UPA suffers from evident coverage non-uniformity.
With omnidirectional antennas, the UPA exhibits strong periodic beam fluctuations, while with directional antennas, high sensing SNR is mainly concentrated around the boresight direction, leaving severe low SNR regions at large angular deviations.
In contrast, the spherical DCAA provides much more uniform and higher sensing SNR over the considered angular region.
Moreover, as each sUPA only needs to cover a small angular sector, when directional antennas are employed, the higher sensing SNR coverage can be achieved, which demonstrates the superiority of the spherical DCAA in terms of the 3D sensing coverage than conventional UPA.

The above results are based on point-wise sensing SNR comparison.
In the next subsection, we further analyze the average probability of detection over the prescribed sensing region, that provides a more direct characterization of the sensing coverage advantage brought by the spherical DCAA.

\subsection{Average Probability of Detection}

For a non-fluctuating point target, consider the linear detection under the circularly symmetric complex Gaussian (CSCG) noise, the probability of target detection with a given false alarm probability $P_{\rm fa}$ is given by \cite{xiao2022waveform}
\begin{equation}\label{PD}
P_{\rm D}(\phi_s,\theta_s)
=
Q_1\left(
\sqrt{2\gamma(\phi_s,\theta_s)},
\sqrt{-2\ln P_{\rm fa}}
\right),
\end{equation}
where $Q_1(\cdot,\cdot)$ denotes the first-order Marcum $Q$-function.

Assuming that the target angle is uniformly distributed over
the angular sensing region
$\Omega\triangleq[\phi_{\min},\phi_{\max}]\times[\theta_{\min},\theta_{\max}]$,
the average probability of detection is defined as
\begin{equation}\label{avg_PD_exact}
\begin{aligned}
&\bar P_{\rm D}
=
\frac{1}{|\Omega|}
\int_{\theta_{\min}}^{\theta_{\max}}
\int_{\phi_{\min}}^{\phi_{\max}}
P_{\rm D}(\phi_s,\theta_s)
d\phi_s d\theta_s\\
&=\frac{1}{|\Omega|}
\int_{\theta_{\min}}^{\theta_{\max}}
\int_{\phi_{\min}}^{\phi_{\max}}
Q_1\left(
\sqrt{2\gamma(\phi_s,\theta_s)},
\sqrt{-2\ln P_{\rm fa}}
\right)
d\phi_s d\theta_s.
\end{aligned}
\end{equation}
Note that it is difficult to directly obtain the closed-form expression of \eqref{avg_PD_exact} due to the nonlinear Marcum-$Q$ function and the angle-dependent sensing SNR.

Nevertheless, as $Q_1(\sqrt{2\gamma(\phi_s,\theta_s)},\sqrt{2\ln P_{\rm fa}})$ is monotonically increasing with the sensing SNR $\gamma(\phi_s,\theta_s)$ for a given $P_{\rm fa}$, we obtain that if
\begin{equation}\label{SNR compare}
\gamma_s(\phi_s,\theta_s)\ge \gamma_{\rm UPA}(\phi_s,\theta_s), \forall (\phi_s,\theta_s)\in\Omega,
\end{equation}
it follows that
$\bar{P}_{D,s}\ge \bar{P}_{D,\rm UPA}$,
where $\bar{P}_{s,D}$ and $\bar{P}_{s,\rm UPA}$ are the average probability of detection of the spherical DCAA and the conventional UPA, respectively, derived according to \eqref{PD} and \eqref{avg_PD_exact}.
In particular, according to the sensing SNR analysis in Section~\ref{sensing SNR analysis}, \eqref{SNR compare} is guaranteed if
\begin{equation}\label{SNR compare 2}
\gamma_{s,\text{worst}}\ge\gamma_{\text{UPA,best}}(\phi_s,\theta_s), \forall(\phi_s,\theta_s)\in\Omega.
\end{equation}
By comparing \eqref{worst sensing SNR} and \eqref{best case UPA}, it has that \eqref{SNR compare 2} holds as long as
\begin{equation}
G_{n^*}\left(\frac{\Delta\phi(\vartheta_{n^*})}{2},\frac{\Delta\theta}{2}\right)\ge\frac{G_{\text{UPA}}(\phi_s,\theta_s)}{|f(\phi_s,\theta_s;\eta_{n^*},\vartheta_{n^*})|^2}.
\end{equation}
This indicates that the sensing SNR of the spherical DCAA is consistently superior to that of UPA, achieving a higher average
probability of detection.

\begin{figure}
    \centering
    {
    \includegraphics[width=0.48\textwidth]{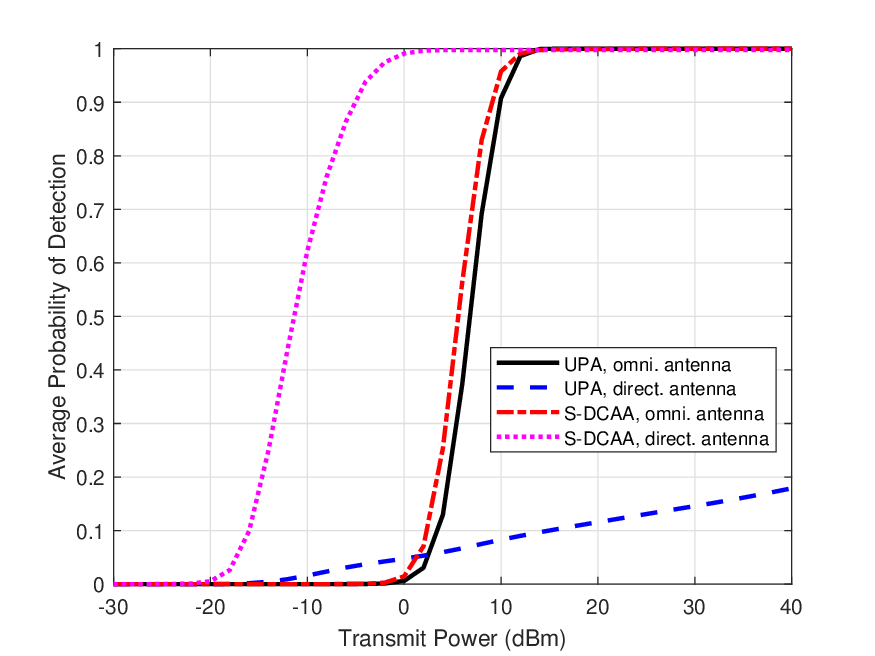}
    }
  \caption{Average probability of detection comparison for the spherical DCAA and the conventional UPA with omnidirectional and directional antennas.}\label{Pd compare}\vspace{-15pt}
\end{figure}
Fig.~\ref{Pd compare} compares the average probability of detection
between the spherical DCAA and the conventional UPA with
omnidirectional and directional antenna patterns, respectively.
It is expected that the spherical DCAA with directional antennas
achieves the best detection performance among all considered schemes.
Moreover, even with omnidirectional antennas, the spherical DCAA
still achieves slightly better detection performance than the UPA.
This is because that the spherical DCAA can effectively mitigate
the SNR degradation at large azimuth and elevation angles, as
shown in Fig.~\ref{Sensing SNR comparison}(c).
In contrast, although the UPA with directional antennas can improve
the peak sensing SNR near the array boresight, its fixed orientation
causes severe SNR degradation when the target direction deviates from
the boresight direction, resulting in the worst average detection
performance.

\subsection{CRLB for Angle Estimation}
In this subsection, we further analyze the CRLBs of the spherical DCAA for both azimuth and elevation angle estimation.
According to \eqref{sensing rx}, the received sensing signal corresponding to the $n$th probing beam is
\begin{equation}
y_n = \alpha_s r_n(\phi_s,\theta_s) + z_n, n=1,\cdots, N,
\end{equation}
where $z_n\sim\mathcal{CN}(0,\sigma_s^2)$ and $r_n(\phi_s,\theta_s)$ is the sensing response function, which is given by
\begin{equation}\label{sensing response}
\small
r_n(\phi_s,\theta_s)=e^{j\omega_0}\sqrt{M^2G(\eta_n-\phi_s,\vartheta_n-\theta_s)}\mathcal{H}_M(x_{1,n})\mathcal{H}_M(x_{2,n}),
\end{equation}
according to \eqref{array response}, with
\begin{equation}
\left\{\begin{aligned}
&x_{1,n}\triangleq\cos\theta_s\sin(\phi_s-\eta_n)\\ &x_{2,n}\triangleq\sin\theta_s\cos\vartheta_n-\cos\theta_s\sin\vartheta_n\cos(\phi_s-\eta_n).
\end{aligned}\right.
\end{equation}
Note that since the target reflection coefficient $\alpha_s$ is independent to the angle estimation, we assume that it is a deterministic coefficient during the observation interval~\cite{wang2024cramer}.

Define the stacked observation vector as $\mathbf{y}=[y_1,\cdots,y_N]^T$.
The mean of $\mathbf y$ is given by $\boldsymbol{\mu}(\phi_s,\theta_s)=\alpha_s \mathbf{r}(\phi_s,\theta_s)$, where
$\mathbf{r}(\phi_s,\theta_s)=[r_1(\phi_s,\theta_s),\cdots,r_N(\phi_s,\theta_s)]^T$ is the sensing response vector.
Let the parameter vector of interest be $\boldsymbol \xi = [\phi_s,\theta_s]^T$.
The Fisher information matrix (FIM) w.r.t. $\boldsymbol \xi$ is
\begin{equation}
\begin{aligned}
\mathbf{J}
&= \frac{2|\alpha_s|^2}{\sigma_s^2}
\Re
\left\{
\begin{bmatrix}
\mathbf r_\phi^H\mathbf r_\phi
&
\mathbf r_\phi^H\mathbf r_\theta
\\
\mathbf r_\theta^H\mathbf r_\phi
&
\mathbf r_\theta^H\mathbf r_\theta
\end{bmatrix}
\right\}\triangleq
\begin{bmatrix}
J_{\phi\phi} & J_{\phi\theta}\\
J_{\theta\phi} & J_{\theta\theta}
\end{bmatrix},
\end{aligned}
\end{equation}
where $\mathbf r_\phi\triangleq\frac{\partial\mathbf r(\phi_s,\theta_s)}{\partial\phi_s}$ and $
\mathbf r_\theta\triangleq\frac{\partial\mathbf r(\phi_s,\theta_s)}{\partial\theta_s}$.
Therefore, the CRLBs of azimuth and elevation angle estimation are

\begin{subequations}\label{CRLB}
\begin{equation}
{\rm CRLB}_{\phi_s} = \frac{J_{\theta\theta}}
{J_{\phi\phi}J_{\theta\theta}-|J_{\phi\theta}|^2},
\end{equation}
\begin{equation}
{\rm CRLB}_{\theta_s}
=
\frac{J_{\phi\phi}}
{J_{\phi\phi}J_{\theta\theta}-|J_{\phi\theta}|^2}.
\end{equation}
\end{subequations}

To gain some insights, suppose that the target direction is aligned with the orientation of the $n^*$th sUPA, i.e., $\phi_s\approx \eta_{n^*}$ and $\theta_s\approx\vartheta_{n^*}$.
Then, we have the following Theorem 1.

{\it Theorem 1}: Under the assumption of the slow variation of antenna pattern, i.e., $G(\eta_n-\phi_s,\vartheta_n-\theta_s)\approx G(0,0)$, $\forall n$, the CRLBs for azimuth and elevation angle estimation can be approximately expressed as
\begin{subequations}\label{CRLB approx}
\begin{equation}
  {\rm CRLB}_{\phi_s}\approx \frac{2\sigma_s^2}{|\alpha_s|^2\pi^2(M-1)^2M^2G(0,0)\cos^2\theta_s},
\end{equation}
\begin{equation}
  {\rm CRLB}_{\theta_s}\approx \frac{2\sigma_s^2}{|\alpha_s|^2\pi^2(M-1)^2M^2G(0,0)}.
\end{equation}
\end{subequations}
\begin{IEEEproof}
Please refer to Appendix.
\end{IEEEproof}

By contrast, according to \eqref{UPA response} and \eqref{KPC}, for the conventional UPA, the received sensing signal corresponding to the $(p,q)$th KPC probing beam is
\begin{equation}
y_{p,q} = \alpha_s r_{\rm UPA}(\phi_s,\theta_s;p,q) + z_{p,q},
\end{equation}
where $z_{p,q}\sim\mathcal{CN}(0,\sigma_s^2)$ and
$r_{\rm UPA}(\phi_s,\theta_s;p,q)$ denotes the sensing response of the
UPA. According to the UPA response and KPC beamforming model, we have
\begin{equation}
r_{\rm UPA}(\phi_s,\theta_s;p,q)
=
e^{j\omega_0}
\sqrt{
M^2G(0,0)
}
\mathcal H_M(x_h)
\mathcal H_M(x_v),
\end{equation}
with
$x_h\triangleq\cos\theta_s\sin\phi_s-\frac{2p}{M}$
and
$x_v\triangleq\sin\theta_s-\frac{2q}{M}$.
Here, for fairness comparison, we consider an omnidirectional antenna pattern, i.e., $G(\phi_s,\theta_s)\approx G(0,0)$.
%Define the stacked response vector over all KPC probing beams as
%\begin{equation}
%\small
%\mathbf r_{\rm UPA}(\phi_s,\theta_s)
%=
%[
%r_{\rm UPA}(\phi_s,\theta_s;p_1,q_1),
%\cdots,
%r_{\rm UPA}(\phi_s,\theta_s;p_{N_b},q_{N_b})
%]^T,
%\end{equation}
%where $N_b=N_hN_v$ denotes the number of probing beams.
%Then, the FIM w.r.t. $\xi=[\phi_s,\theta_s]^T$ is given by
%\begin{equation}
%\mathbf J_{\rm UPA}
%=
%\frac{2|\alpha_s|^2}{\sigma_s^2}
%\Re
%\left\{
%\begin{bmatrix}
%\mathbf r_{{\rm UPA},\phi}^H
%\mathbf r_{{\rm UPA},\phi}
%&
%\mathbf r_{{\rm UPA},\phi}^H
%\mathbf r_{{\rm UPA},\theta}
%\\
%\mathbf r_{{\rm UPA},\theta}^H
%\mathbf r_{{\rm UPA},\phi}
%&
%\mathbf r_{{\rm UPA},\theta}^H
%\mathbf r_{{\rm UPA},\theta}
%\end{bmatrix}
%\right\},
%\end{equation}
%where
%$\mathbf r_{{\rm UPA},\phi}=\frac{\partial\mathbf r_{\rm UPA}(\phi_s,\theta_s)}{\partial\phi_s}$ and $
%\mathbf r_{{\rm UPA},\theta}=\frac{\partial\mathbf r_{\rm UPA}(\phi_s,\theta_s)}{\partial\theta_s}.
%$
%The CRLBs of the UPA for azimuth and elevation angle estimation are respectively given by
%\begin{equation}\label{CRLB UPA}
%{\rm CRLB}_{\phi_s}^{\rm UPA}
%=
%\mathbf J_{\rm UPA}^{-1}(1,1),
%\quad
%{\rm CRLB}_{\theta_s}^{\rm UPA}
%=
%\mathbf J_{\rm UPA}^{-1}(2,2).
%\end{equation}
Therefore, similar to \eqref{CRLB approx}, suppose that the $(p^*,q^*)$th KPC beam satisfies
$\cos\theta_s\sin\phi_s-\frac{2p^\star}{M}
\approx0$
and
$\sin\theta_s-\frac{2q^\star}{M}
\approx0$.
The CRLBs of conventional UPA for elevation and azimuth angle estimation can be approximated as
\begin{subequations}
\begin{equation}
{
{\rm CRLB}_{\phi_s}^{\rm UPA}
\approx
\frac{
2\sigma_s^2
}{
|\alpha_s|^2
\pi^2(M-1)^2
M^2
G(0,0)
\cos^2\theta_s
\cos^2\phi_s
}
}
\end{equation}
\begin{equation}
{
{\rm CRLB}_{\theta_s}^{\rm UPA}
\approx
\frac{
2\sigma_s^2
}{
|\alpha_s|^2
\pi^2(M-1)^2
M^2
G(0,0)
\cos^2\theta_s
}
}.
\end{equation}
\end{subequations}

Note that compared with the spherical DCAA, the conventional UPA suffers from the direction-dependent performance losses, where the azimuth and elevation CRLBs are degraded by the factors $\cos^2\phi_s$ and $\cos^2\theta_s$, respectively.
Since the effective aperture of the UPA decreases significantly when the target direction deviates from the array broadside direction.

\begin{figure}
    \centering
        \subfigure[CRLB for azimuth angle estimation, $\phi_s=80^{\circ}$.]{
    \includegraphics[width=0.48\textwidth]{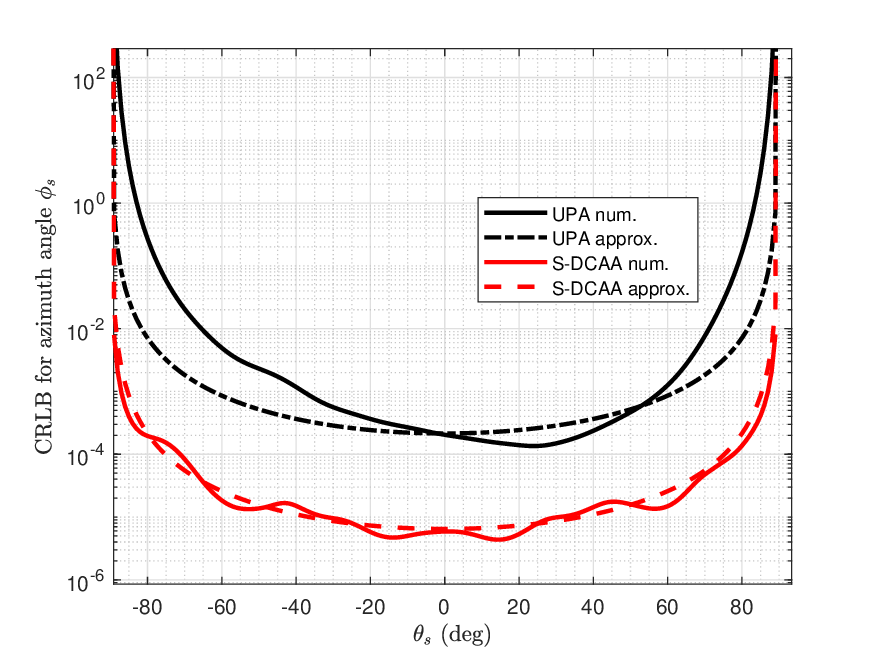}
    }
    \hspace{-15pt}
    \subfigure[CRLB for elevation angle estimation.]{
    \includegraphics[width=0.48\textwidth]{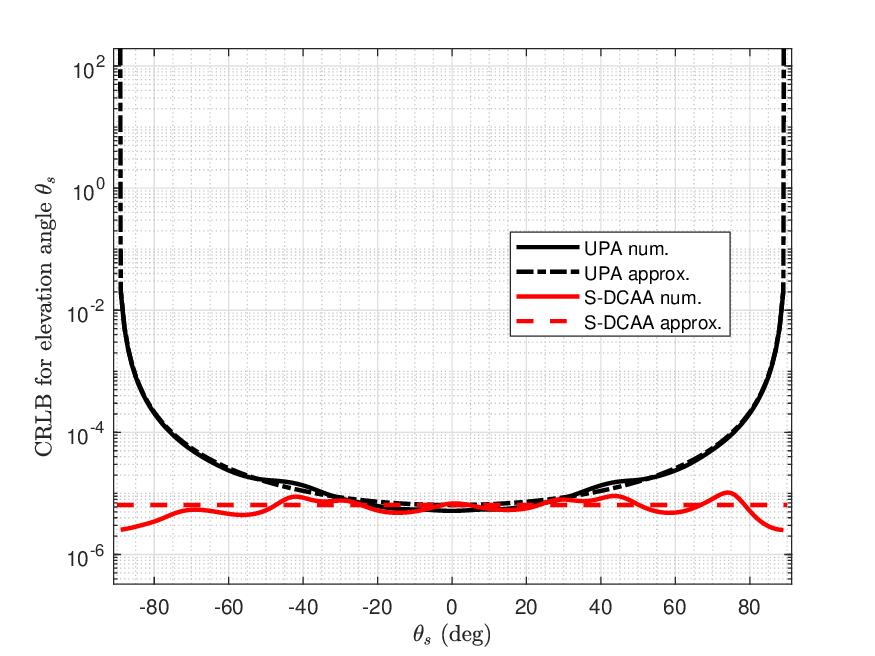}
    }
  \caption{CRLB comparison of spherical DCAA and conventional UPA.}\label{CRLB results}\vspace{-15pt}
\end{figure}
Fig.~\ref{CRLB results} compares the CRLB performance of the spherical DCAA and the conventional UPA for azimuth and elevation angle estimation, respectively.
We obtain that the spherical DCAA achieves substantially lower CRLBs over the entire angular region and maintains relatively stable estimation accuracy due to its 3D spherical deployment.
By contrast, the CRLBs of the UPA increase rapidly in large angle directions due to the antenna aperture projection loss.
Moreover, the approximate CRLBs closely match with the exact results near the boresight regions, which verifies the accuracy of the proposed analytical approximation.

The aforementioned sensing performance is critically determined by the sensing SNR.
Moreover, from \eqref{communication SINR} and \eqref{sensing_SNR}, it observes that for the spherical DCAA, both communication SINR and sensing SNR depend on the array selection matrix $\mathbf{U}$ and the digital beamforming matrix $\mathbf{F}$.
Therefore, in the following, we investigate a joint array selection and beamforming optimization problem for low-altitude ISAC.

\section{Joint Array Selection and Beamforming Optimization for ISAC}\label{optimization problem}
To guarantee the sensing coverage over the prescribed region $\mathcal{A}$, we aim to maximize the worst-case sensing SNR, while satisfying the communication SINR requirement.
The corresponding optimization problem is formulated as
\begin{align}
({\rm P0}):\max\limits_{\mathbf{U},\mathbf{F}}\min\limits_{\mathbf{t}\in\mathcal{A}}\quad &\gamma_{s}(\mathbf{U},\mathbf{F};\mathbf{t})\label{P0}\\
{\it s.t.}\quad&\gamma_{k}(\mathbf{U},\mathbf{F})\ge \Gamma_k, k\in\mathcal{K},\tag{\ref{P0}{a}}\label{P0_a}\\
& \sum\nolimits_{k=1}^{K}\|\mathbf f_k\|^2\le P_t, \tag{\ref{P0}{b}}\label{P0_b}\\
&     \mathbf U\in\{0,1\}^{N_{\rm RF}\times N}, \tag{\ref{P0}{c}}\label{P0_c}\\
&     \|\mathbf U_{i,:}\|_0=1, i=1,\cdots,N_{\rm RF},\tag{\ref{P0}{d}}\label{P0_d}\\
&     \|\mathbf U_{:,j}\|_0\le1, n=1,\cdots,N, \tag{\ref{P0}{e}}\label{P0_e}
\end{align}
where $\Gamma_{k}>0$ denotes the communication SINR threshold of the $k$th UE.
Note that $\rm (P0)$ is a mixed-integer non-convex optimization
problem.
To gain useful insights, we first study a special case of orthogonal multi-user communication and point-target sensing, and then extend the results to the general multi-user communication and regional sensing case.

\subsection{Orthogonal Multi-User Communication and Point Sensing}\label{special case}
According to \eqref{communication SINR} and \eqref{sensing_SNR}, denote by $\mathbf{a}_s(\mathbf{U})\triangleq\alpha_s\mathbf{U}\mathbf{r}(\mathbf{w}_s)$ and $\tilde{\mathbf{h}}_{k}(\mathbf{U})\triangleq\mathbf{U}\bar{\mathbf{h}}_{ek}$ the effective sensing and communication channel of the $k$th user, respectively, under a given array selection matrix $\mathbf{U}$.
Note that due to the energy-focusing capability of the spherical DCAA, when $M\gg KL$ and $N_{\rm RF} > KL$, we can assume that the effective communication channels of difference users as well as the sensing channel are asymptotically orthogonal, i.e., $\tilde{\mathbf{h}}_{k1}^H(\mathbf{U})\tilde{\mathbf{h}}_{k2}(\mathbf{U})=0$, $k1\neq k2$, and $\mathbf{a}_s^H(\mathbf{U})\tilde{\mathbf{h}}_k(\mathbf{U})=0$, $k\in\mathcal{K}$.

In this case,  the sensing and communication SNR can be rewritten as
\begin{equation}\label{orthogonal SNR}
\begin{aligned}
&\gamma_s(\mathbf{U},\mathbf{F};\mathbf{t}_s) = \sum\nolimits_{k=1}^{K}\left|\mathbf{a}_s(\mathbf{U})^H\mathbf{f}_k\right|^2/\sigma_s^2\\
&\gamma_k(\mathbf{U},\mathbf{f}_k) = |\tilde{\mathbf h}_{k}^H(\mathbf U)\mathbf f_k|^2/\sigma^2
\end{aligned}
\end{equation}
Thus, the problem $(\rm P0)$ reduces to
\begin{equation}\label{problem 1}
\begin{aligned}
({\rm P1}):\max\limits_{\mathbf{U},\mathbf{F}}\quad &\sum\nolimits_{k=1}^{K}\left|\mathbf{a}_s(\mathbf{U})^H\mathbf{f}_k\right|^2\\
{\it s.t.}\quad&|\tilde{\mathbf h}_{k}^H(\mathbf U)\mathbf f_k|^2\ge \Gamma_k\sigma^2, k\in\mathcal{K},\\
& \eqref{P0_b}\text{-}\eqref{P0_e}.
\end{aligned}
\end{equation}

To solve this problem, first consider a fixed $\mathbf{U}$ and the beamforming subproblem of $(\rm P1)$ is formulated as
\begin{equation}
\begin{aligned}
({\rm P1.1}):\quad
\max_{\mathbf F}\quad
&
\sum\nolimits_{k=1}^{K}\left|\mathbf{a}_s(\mathbf{U})^H\mathbf{f}_k\right|^2
\\
{\rm s.t.}\quad
&
|\tilde{\mathbf{h}}_{k}^H(\mathbf U)\mathbf f_k|^2
\ge
\Gamma_k\sigma^2,
\quad k\in\mathcal K,
\\
&
\sum\nolimits_{k=1}^{K}\|\mathbf f_k\|^2
\le
P_t.
\end{aligned}
\label{P1_multi}
\end{equation}
Note that under the orthogonality assumption, the optimal beamformer vector of each user $k$ can be decomposed as
\begin{equation}\label{f_special}
\mathbf{f}_k^* = \sqrt{P_{c,k}} \mathbf{e}_k + \sqrt{P_{s,k}}\mathbf{e}_s, k\in\mathcal{K},
\end{equation}
where $\mathbf{e}_k = \tilde{\mathbf{h}}_k(\mathbf{U})/\|\tilde{\mathbf{h}}_k(\mathbf{U})\|$, $\mathbf{e}_s=\mathbf{a}_s(\mathbf{U})/\|\mathbf{a}_s(\mathbf{U})\|$, and the communication power required by user $k$ is
$
P_{c,k} = \frac{\Gamma_k\sigma^2}{\|\tilde{\mathbf{h}}_k(\mathbf{U})\|^2}.
$
To make the problem $(\rm P1.1)$ feasible, the communication threshold should satisfy that
$\sum\nolimits_{k=1}^{K}\frac{
\Gamma_k\sigma^2
}{
\|\tilde{\mathbf{h}}_{k}(\mathbf U)\|^2
}
\le
P_t$.

The remaining power after satisfying all communication constraints is
$P_s=P_t-\sum\nolimits_{k=1}^{K}P_{c,k}$.
By evenly allocating the sensing power among $K$ beamformers with $P_{s,k}=P_s/K$, the closed-form optimal beamformer is
\begin{equation}
\begin{aligned}
\mathbf f_k^\star
=&
\sqrt{
\frac{
\Gamma_k\sigma^2
}{
\|\tilde{\mathbf h}_{k}(\mathbf U)\|^2
}
}
\mathbf{e}_k+
\sqrt{
\frac{
P_t-
\sum_{k=1}^{K}
\frac{
\Gamma_k\sigma^2
}{
\|\tilde{\mathbf h}_{k}(\mathbf U)\|^2
}
}{K}
}\mathbf{e}_s, k\in\mathcal{K}.
\end{aligned}
\label{f_star}
\end{equation}

By substituting \eqref{f_star} into \eqref{orthogonal SNR}, the corresponding maximum sensing SNR is
\begin{equation}
\gamma_s^\star(\mathbf U)
=
\frac{
K\|\mathbf a_s(\mathbf U)\|^2
}{
\sigma_s^2
}
\left(
P_t
-
\sum\nolimits_{k=1}^{K}
\frac{
\Gamma_k\sigma^2
}{
\|\tilde{\mathbf{h}}_{k}(\mathbf U)\|^2
}
\right).
\label{gamma_multi_closed}
\end{equation}
Then, the problem $(\rm P1)$ can reduce to
\begin{equation}
\begin{aligned}
({\rm P1.2}):\max\limits_{\mathbf{U}}\quad &\|\mathbf{a}_s(\mathbf{U})\|^2\left(
P_t
-
\sum\nolimits_{k=1}^{K}
\frac{
\Gamma_k\sigma^2
}{
\|\tilde{\mathbf{h}}_{k}(\mathbf U)\|^2
}
\right)\\
{\it s.t.}\quad
& \eqref{P0_c} \text{-} \eqref{P0_e}.
\end{aligned}
\end{equation}
This is a combinatorial problem.
Although  exhaustive search can obtain the optimal solution, it has prohibitive complexity.
By exploiting the closed-form sensing SNR in \eqref{gamma_multi_closed}, we propose a low-complexity greedy-based sUPA selection method.

Specifically, for the preselected sUPA index set $\mathcal{S}$, define the selection metric as
\begin{equation}
\zeta(\mathcal S)
=
A_s(\mathcal S)
\left(
P_t
-
\sum\nolimits_{k=1}^{K}
\frac{
\Gamma_k\sigma^2
}{
H_k(\mathcal S)
}
\right),
\label{selection_metric_multi_orthogonal}
\end{equation}
where $A_s(\mathcal{S})$ and $H_k(\mathcal{S})$ are given by
\begin{equation}
A_s(\mathcal{S})=\sum\nolimits_{n\in\mathcal{S}}|a_{s,n}|^2, \quad H_k(\mathcal{S})=\sum\nolimits_{n\in\mathcal{S}}|h_{k,n}|^2
\end{equation}
with $a_{s,n}$ and $h_{c,k,n}$ representing the $n$th entries of the sensing
and communication channels before selection, respectively.
Here,  it comments that the selected sUPA index set should satisfy
$
P_t
-
\sum\nolimits_{k=1}^{K}
\frac{
\Gamma_k\sigma^2
}{
H_k(\mathcal S)
}
\ge 0.
$

Based on \eqref{selection_metric_multi_orthogonal}, a low-complexity
greedy-based array selection method can be developed.
To make the orthogonality assumption valid, the initial selected set $\mathcal{S}_0$ could include the index of the sUPAs matched to the dominant communication and sensing directions, which is given by $\mathcal{S}^{(0)}=\{n_s,n_1^*,\cdots,n_K^*\}$, where
\begin{equation}
\begin{aligned}
&n_k^* = \arg\max\limits_{n\in\mathcal{N}}|r_n(\mathbf{w}_k)|^2, k\in\mathcal{K},\\
&n_s = \arg\max\limits_{n\in\mathcal{N}}|r_n(\mathbf{w}_s)|^2.
\end{aligned}
\end{equation}

At each iteration step $i$, the remain sUPA that gives
the largest increase of $\zeta(\mathcal S)$ is selected, i.e.,
\begin{equation}
n^\star
=
\arg\max_{n\in\mathcal N\setminus\mathcal S}
\zeta(\mathcal S\cup\{n\}).
\end{equation}
Then, the selected index set is updated as
$
\mathcal S^{(i+1)}
=
\mathcal S^{(i)}\cup\{n^\star\}.
$
This procedure is repeated until $|\mathcal S|=N_{\rm RF}$.

\subsection{General Multi-User Communication and Regional Sensing}\label{general case}
For the genercal case, to make the original problem $\rm(P0)$ tractable, the sensing region $\mathcal{A}$ is discretized into $Q$ grid points, denoted by $\mathcal{Q}=\{1,\cdots, Q\}$.
According to \eqref{sensing_SNR}, the received sensing SNR at $\mathbf{t}_q\in\mathcal{A}$ can be rewritten as
\begin{equation}\label{worst case sensing SNR}
\gamma_s(\mathbf{U},\mathbf{F};q) =\kappa_q\sum\nolimits_{k=1}^K\left|\left(\mathbf{U}\mathbf{r}(\mathbf{w}_q)\right)^H\mathbf{f}_k\right|^2,  q\in\mathcal{Q},
\end{equation}
where $\kappa_q=|\alpha_s(\mathbf{t}_q)|^2/\sigma_s^2$.
By introducing an auxiliary variable $\Gamma_s$ that denotes the minimum
sensing SNR over all sensing grid points, $\rm (P0)$ can be reformulated as
\begin{equation}
\begin{aligned}
(\rm P2):\quad
\max_{\mathbf U,\mathbf F,\Gamma_s}\quad
&
\Gamma_s
\\
{\rm s.t.}\quad
&
\gamma_s(\mathbf U,\mathbf F;q)
\ge
\Gamma_s,
\quad
q\in\mathcal Q,
\\
&\eqref{P0_a}\text{-}\eqref{P0_e}.
\end{aligned}
\label{P1_general}
\end{equation}

It is difficult to directly solving this problem, due to the coupling between the optimization variables $\mathbf{U}$ and $\mathbf{F}$, as well as the associated non-convex  constraints.
Motivated by the analytic results for the special case in Section~\ref{special case}, in the following, we propose a novel greedy-based optimization framework  to effectively solve this problem.

\subsubsection{Greedy-Based Optimization Framework}
Starting from the initial set $\mathcal{S}^{(0)}=\{n_1^*,\cdots,n_K^*\}$, at the $u$th iteration $(u\ge 0)$, for the $n$th candidate sUPA, with $n\in\mathcal{N}\setminus\mathcal{S}^{(u)}$, according to \eqref{P0_a} and \eqref{worst case sensing SNR}, the communication feasibility margin and the worst-case sensing SNR can be measured as
\begin{subequations}
\begin{equation}\label{communication feasibility}
  \mu_n = \min\limits_{k\in\mathcal{K}}\frac{\gamma_k(\mathbf{U}_n^{(u)},\mathbf{F}_n^{(u)})}{\Gamma_k},
\end{equation}
\begin{equation}\label{sensing SNR metric}
\Gamma_{s,n}
=
\min_{q\in\mathcal Q}
\gamma_s
(
\mathbf U_n^{(u)},
\mathbf F_n^{(u)};
q
),
\end{equation}
\end{subequations}
where $\mathbf{U}_n^{(u)}={\rm Matrix}\left(S^{(u)}\cup \{n\}\right)$ and $\mathbf{F}_n^{(u)}$ is the solved beamforming matrix under $\mathbf{U}_n^{(u)}$, which will be presented later in the subsequent subsection.

At each iteration $u$, the candidate sUPA  $n$ that maximizes the worst-case sensing SNR while
satisfying the communication constraints is selected, i.e.,
\begin{equation}
n^\star
=
\arg\max_{n\in\mathcal F_{\mathcal S}}
\Gamma_{s,n},
\end{equation}
where
$
\mathcal F_{\mathcal S}
=
\{
n
\mid
\mu_n\ge 1
\}.
$
Note that if no feasible candidate exists, i.e., $\mathcal{F_S}=\emptyset$,
the candidate sUPA that maximizes the communication margin is selected, i.e.,
\begin{equation}
n^\star
=
\arg\max_n
\mu_n.
\end{equation}
Then, the selected set is updated as $\mathcal{S}^{(u+1)}=\mathcal{S}^{(u)}\cup \{n^*\}$.
This procedure is repeated utill $|\mathcal{S}|=N_{\rm RF}$.
The selection matrix can be finally obtained as $\mathbf{U}^*={\rm Matrix}(\mathcal{S}^*)$, with $\mathcal{S}$ being the final selected index set.
In the following, we present that how to optimize the digital beamforming matrix $\mathbf{F}$ with the given selection matrix $\mathbf{U}$ at each iteration $u$.

\subsubsection{Optimization of  $\mathbf{F}$  With Given $\mathbf{U}$}
Define the effective sensing
channels at the $q$th grid as $\tilde{\mathbf r}_q=\mathbf U\mathbf r(\mathbf w_q)$.
The subproblem of $\rm(P2)$ for $\mathbf{F}$ optimization can be formulated as
\begin{equation}
\begin{aligned}
\rm (P2.1):\quad
\max_{\mathbf F,\Gamma_s}\quad
&\Gamma_s\\
{\rm s.t.}\quad
&
\sum\nolimits_{k=1}^{K}
|\tilde{\mathbf r}_q^H\mathbf f_k|^2
\ge
\Gamma_s/\kappa_q,\quad q\in\mathcal Q,\\
&
\frac{|\tilde{\mathbf h}_k^H\mathbf f_k|^2}
{\sum_{i\neq k}|\tilde{\mathbf h}_k^H\mathbf f_i|^2+\sigma^2}
\ge \Gamma_k,\quad k\in\mathcal K,\\
&
\sum\nolimits_{k=1}^{K}\|\mathbf f_k\|^2\le P_t,
\end{aligned}
\end{equation}
where  the communication SINR constraint can be rewritten as the convex second-order cone (SOC) constraints as
\begin{equation}
\begin{aligned}
&\left\|
\tilde{\mathbf h}_k^H\mathbf f_1,\cdots,
\tilde{\mathbf h}_k^H\mathbf f_K,\sigma
\right\|
\le
\sqrt{1+\frac{1}{\Gamma_k}}
{\rm Re}(\tilde{\mathbf h}_k^H\mathbf f_k),\\
&{\rm Im}(\tilde{\mathbf{h}}_k^H\mathbf{f}_k)=0.
\end{aligned}
\label{SOC_comm}
\end{equation}
However, as the sensing constraint is non-convex and difficult be transformed into a SOC constraint, here the successive convex approximation (SCA) technique is adopted to deal with the non-convex sensing constraint.

Specifically, let $g_q(\mathbf{F})=\sum\nolimits_{k=1}^{K}|\tilde{\mathbf{r}}_q^H\mathbf{f}_k|^2$.
Since $g_q(\mathbf{F})$ is convex, it can be lower bounded by its first-order Taylor expansion at $\mathbf{F}^{(i)}$ for the $i$th iteration, i.e.,
\begin{equation}
\begin{aligned}
g_q(\mathbf F)
\ge
&
g_q(\mathbf F^{(i)})
+
2\sum\nolimits_{k=1}^{K}
{\rm Re}\left\{
(\mathbf f_k^{(i)})^H
\tilde{\mathbf r}_q\tilde{\mathbf r}_q^H
(\mathbf f_k-\mathbf f_k^{(i)})
\right\}\\
\triangleq&
\varphi_q(\mathbf F,\mathbf F^{(i)}).
\end{aligned}
\label{SCA_sensing}
\end{equation}
For each iteration $i$, $\rm(P2.1)$ can be
approximated as
\begin{equation}
\begin{aligned}
(\rm P2.1.1):\quad
\max_{\mathbf F,\Gamma_s}\quad
&\Gamma_s\\
{\rm s.t.}\quad
&
\varphi_q(\mathbf F,\mathbf F^{(l)})
\ge
\Gamma_s/\kappa_q,\quad q\in\mathcal Q,\\
&
\eqref{SOC_comm},\quad k\in\mathcal K,\\
&
\sum\nolimits_{k=1}^{K}\|\mathbf f_k\|^2\le P_t,
\end{aligned}
\end{equation}
which is a convex SOCP problem.
It can be efficiently solved by using CVX toolbox.
Then, by iteratively updating the local point as $\mathbf F^{(i+1)}=\mathbf F^*$ and iteratively solving a sequence of $\rm (P2.1.1)$, the problem $\rm(P2.1)$ can be effectively solved.

The overall procedure is summarized in Algorithm 1.
The computational complexity is on the order of $O(NN_{\rm RF}I_{\rm SCA}C_{\rm SOCP})$, where $C_{\rm SCOP}$ denotes the complexity for solving the SCOP subproblem $(\rm P2.1.1)$ in each SCA iteration, and $I_{\rm SCA}$ represents the number of SCA iterations required for convergence.
Although the proposed algorithm is still computational demanding, it provides a tractable solution to the mixed-integer non-convex problem compared to the exhaustive search method.
In practice, the SCA-based beamforming can be replaced by low-complexity beamforming metric, such as maximum ratio transmission (MRT) or minimum mean-square error (MMSE), to further reduce the complexity but at the cost of some performance loss.

\begin{algorithm}[t]
	\caption{Greedy-Based SCA for Solving $(\rm P2)$}
	\label{alg1}
	\textbf{Input}: $\{\bar{\mathbf h}_{e,k},\Gamma_k\}_{k=1}^{K}$,
    $\{\mathbf r(\mathbf w_q),\kappa_q\}_{q=1}^{Q}$, $P_t$,
    $\epsilon$.

    \textbf{Output}: $\mathbf F^*$, $\mathbf U^*$, $\Gamma_s^*$.

    Initialize $\mathcal{S}^{(0)}=\{n_1^*,\cdots,n_K^*\}$, $u=0$.

	\While{$|\mathcal{S}^{(u)}|<N_{\rm RF}$}
    {
        Set $\mathcal{F_S}=\emptyset$.

        \For{$n\in\mathcal{N}\setminus\mathcal{S}^{(u)}$}
        {
           $\mathbf{U}_n^{(u)}={\rm Matrix}(\mathcal{S}^{(u)}\cup n)$.

           Initialize $\mathbf{F}^{(0)}$, $\Gamma_s^{(0)}$, $i=0$.

           \Repeat{
            $|\Gamma_s^{(i+1)}-\Gamma_s^{(i)}|\le\epsilon$
           }
           {
              Solve $(\rm P2.1.1)$ to obtain $\mathbf{F}^*$ and $\Gamma_s^*$.

              Update $\mathbf{F}^{(i+1)} = \mathbf{F^*}$.

              Update $\Gamma_s^{(i+1)} = \Gamma_s^*$.

              $i = i + 1$;

           }

           Update $\mathbf{F}_n^{(u)} = \mathbf{F}^*$.

           Compute $\mu_n = \min\limits_{k\in\mathcal{K}}\frac{\gamma_k(\mathbf{U}_n^{(u)},\mathbf{F}_n^{(u)})}{\Gamma_k}$.

           Compute $\Gamma_{s,n}=\min\limits_{q\in\mathcal{Q}}\gamma_s(\mathbf{U}_n^{(u)},\mathbf{F}_n^{(u)};q)$.

           \If{
              $\mu_n \ge 1$;
           }{
              $\mathcal{F_S} = \mathcal{F_S}\cup \{n\}$.
           }
        }
      \If{
        $\mathcal{F_S}\neq\emptyset$
      }{
        $n^* = \arg\max\limits_{n\in\mathcal{F_S}}\ \Gamma_{s,n}$
      }
      \Else{
        $n^* = \arg\max\limits_{n\in\mathcal{N}\setminus\mathcal{S}^{(u-1)}}\ \mu_n$.
      }
      Update $\mathcal{S}^{(u+1)} = \mathcal{S}^{(u)}\cup\{n^*\}$.

      $u = u + 1$;
    }

    Output $\mathbf{U}^* ={\rm Matrix}(\mathcal{S}^{(u)})$, $\mathbf{F}^*$, $\Gamma_s^*$.
\end{algorithm}

\section{Simulation Results}\label{simulation results}
In this section, simulation results are presented to evaluate the performance of spherical DCAA for low-altitude ISAC systems.
For spherical DCAA, each sUPA consists of $M\times M = 64$ antenna elements.
The maximum elevation and azimuth orientation angles of the sUPA are $\vartheta_{\max}=\eta_{\max}=\pi/2$.
According to the architecture described in Section~\ref{system model}, the resulting spherical DCAA contains $N = 101$ candidate sUPAs.
For fairness comparison, a conventional UPA composed of $M\times M =64$ antenna elements is adopted as the benchmark, which employs the KPC with the beamforming codewords are given by $\mathbf{f}_{q,p}=\mathbf{f}_q\otimes\mathbf{f}_{p}$, where $q,p\in\{-4,\cdots,-1,0,1,\cdots,4\}$.
For both spherical DCAA and UPA, the number of RF chains is set to $N_{\rm RF} = 7$.

In the following, we first present simulation results for the special case in Section~\ref{special case} to gain insights into the proposed joint array selection and beamforming design.
Then, the effectiveness of the proposed optimization framework is evaluated in the general multi-user communication and regional sensing scenario described in Section~\ref{general case}.

\begin{figure}[t]
    \centering
        \subfigure[Convergence of the proposed greedy sUPA selection method.]{
    \includegraphics[width=0.45\textwidth]{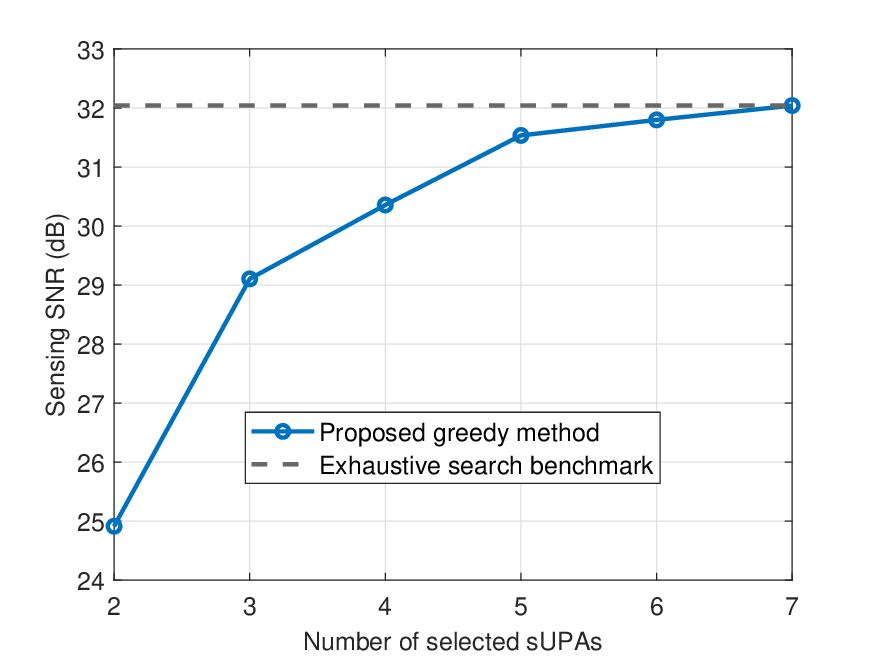}
    }
    \hspace{-15pt}
    \subfigure[Selected sUPAs of the spherical DCAA.]{
    \includegraphics[width=0.45\textwidth]{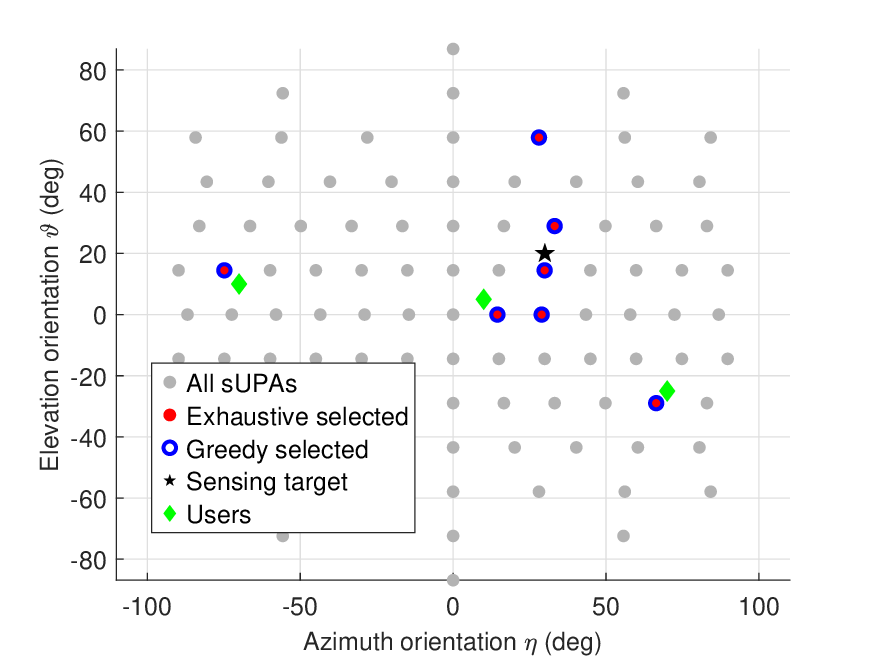}
    }
    \hspace{-15pt}
    \subfigure[Optimized transmit beampattern with selected sUPA]{
    \includegraphics[width=0.45\textwidth]{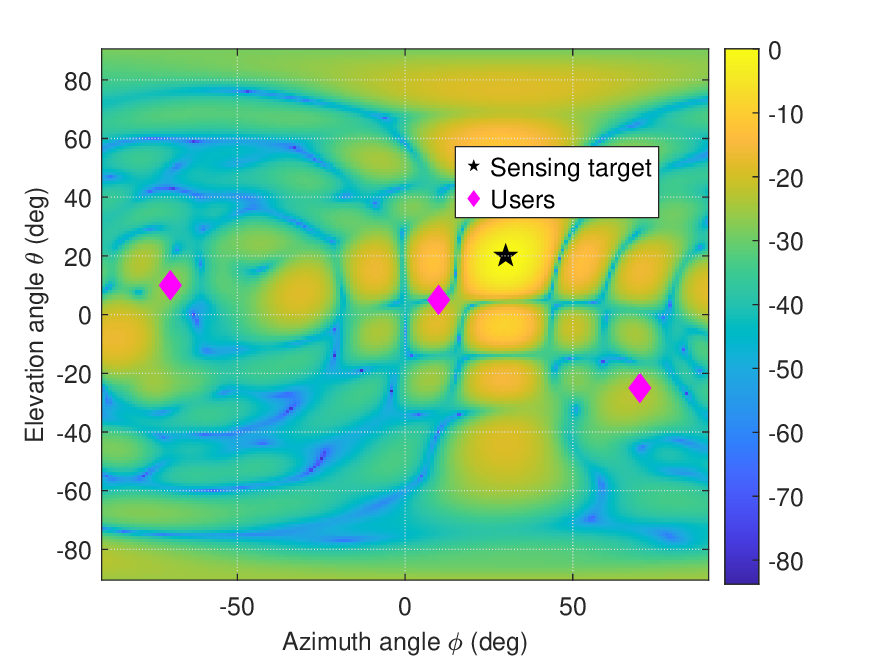}
    }
  \caption{Performance evaluation of the proposed greedy sUPA selection method for the special case.}\label{ISAC_special}\vspace{-15pt}
\end{figure}
\subsection{Orthogonal Multi-User Communication and Point Sensing}
For the special case, we consider three ground communication UEs and a point-like sensing target that are spatially well separated.
As a result, the communication and sensing channels are assumed to be approximately orthogonal.

Fig.~\ref{ISAC_special}(a) illustrates the convergence of the proposed greedy-based sUPA selection algorithm presented in Section~\ref{special case}.
It can be observed that the sensing SNR monotonically increases as more sUPAs are selected.
This is because additional RF chains provide more spatial DoF, which can be exploited to improve the sensing performance once the communication SINR requirements have been satisfied.
Furthermore, the sensing performance achieved by the proposed greedy method approaches to that of exhaustive search method, which demonstrates the effectiveness of the proposed method.

\begin{figure}[t]
    \centering
    {
    \includegraphics[width=0.48\textwidth]{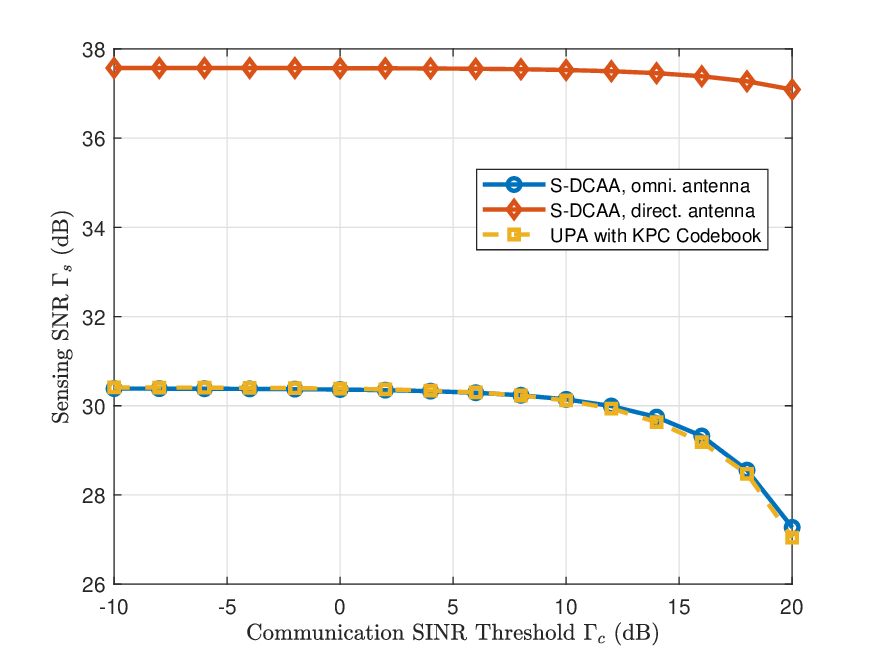}
    }
  \caption{Comparison of the communication-sensing performance tradeoff between the spherical DCAA and UPA for low-altitude ISAC systems.}\label{ISAC_tradeoff_special}\vspace{-15pt}
\end{figure}
Furthermore, the final sUPAs selection results is shown in Fig.~\ref{ISAC_special}(b), where the gray dots denotes the placement of all sUPAs, while the red and blue markers correspond to the sUPAs selected by the exhaustive search and the proposed greedy method, respectively.
Note that the selected sUPAs are mainly distributed around the directions of the sensing target and communication UEs, to provide strong communication and sensing channel gains, thereby improving the sensing SNR while guaranteeing the communication performance.
Moreover, it observes that the sUPAs selected by the proposed greedy method coincide with those obtained by exhaustive search, which validates the effectiveness of the proposed selection criterion based on communication feasibility and sensing performance.
In addition, Fig.~\ref{ISAC_special}(c) depicts the optimized transmit beampattern obtained by using the selected sUPAs.
It can be obtained that the main beam is steered towards the sensing target direction to maximize the sensing SNR, while the sidelobe energy is allocated towards the UE directions to satisfy the communication requirements.

Fig.~\ref{ISAC_tradeoff_special} further compares the communication--sensing trade-off achieved by the spherical DCAA and conventional UPA systems.
It observes that the sensing SNR decreases as the communication SINR threshold $\Gamma_c$ increases.
Moreover, the spherical DCAA equipped with omnidirectional antenna elements exhibits similar performance as that of conventional UPA with KPC for point target sensing.
In contrast, where directional antenna elements are adopted, spherical DCAA achieves a more favorable communication--sensing tradeoff than the conventional UPA.
Moreover, it is worth noting that the performance gain becomes larger as the communication SNR threshold increases.
This attributes to that each sUPA of the spherical DCAA only needs to cover a portion of overall coverage region, highly directional antenna elements with enhanced energy-focusing capability can be employed, leading to the antenna gain for both communication and sensing performance improvement.

\subsection{General Multi-User Communication and Regional Sensing}
We now consider a general low-altitude ISAC scenario, where the BS equipped with a spherical DCAA simultaneously communicates with multiple UEs and provides aerial sensing coverage over a prescribe region.
In contrast to the orthogonal multi-user communication case, inter-user interference (IUI) among different UEs is considered.
The sensing region spans  $[30^\circ,80^\circ]$ in elevation and $[10^\circ,50^\circ]$ in azimuth, respectively.

First, we evaluate the effectiveness of the proposed greedy-based optimization method.
To reduce the computational complexity of the proposed method, as discussed in Section~\ref{general case}, the low-complexity MRT and MMSE beamformimg metrics are also considered and evaluated.
Specifically, let $\mathbf{F}_{\rm aux}=[\mathbf{f}_{1,\rm aux},\cdots,\mathbf{f}_{K,\rm aux}]$ denote the auxiliary beamforming matrix.
Under the assumption that the aerial sensing and ground communication channels are approximately orthogonal, the auxiliary beamforming vector can be constructed as
\begin{equation}
  \mathbf{f}_{k,\rm aux} = \sqrt{\rho_{c,k}}\mathbf{e}_k + \sqrt{\rho_{s,k}}\mathbf{e}_s, \quad k = 1,\cdots, K,
\end{equation}
where $\mathbf{e}_s=\mathbf{a}(\mathbf{U})/\|\mathbf{a}(\mathbf{U})\|$ is the normalized sensing steering vector, $\mathbf{e}_k$ is the normalized communication vector computed according to the chosen metric, e.g., MMSE or MRT, $\rho_{c,k}$ and $\rho_{s,k}$ are the power ratio factor, with $0<\rho_{c,k},\rho_{s,k}<1$.
By substituting $\mathbf{F}_{\rm aux}$ into \eqref{communication feasibility} and \eqref{sensing SNR metric}, the greedy-based array selection is performed following Algorithm~1.

\begin{figure}[t]
    \centering
    {
    \includegraphics[width=0.48\textwidth]{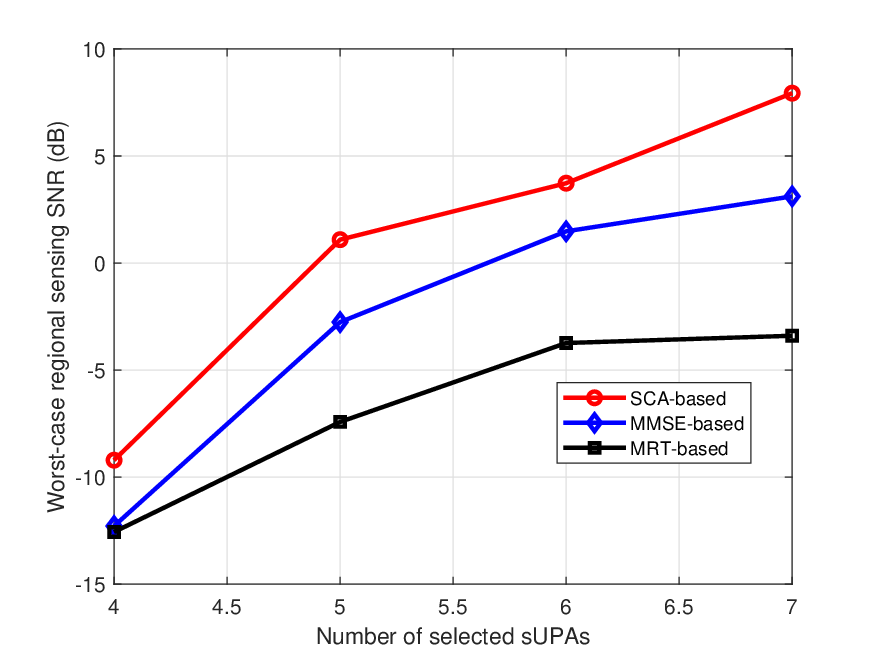}
    }
  \caption{The performance of the proposed greedy selection method under different beamforming schemes}\label{ISAC_general_covergence}\vspace{-15pt}
\end{figure}

Fig.~\ref{ISAC_general_covergence} shows the performance of the proposed greedy-based array selection method under different beamforming metrics.
It can be observed that the SCA-based beamforming outperforms the MMSE- and MRT-based surrogate metrics, even with higher computational complexity, which demonstrates the effectiveness of the proposed method.

\begin{figure}[t]
    \centering
    {
    \includegraphics[width=0.48\textwidth]{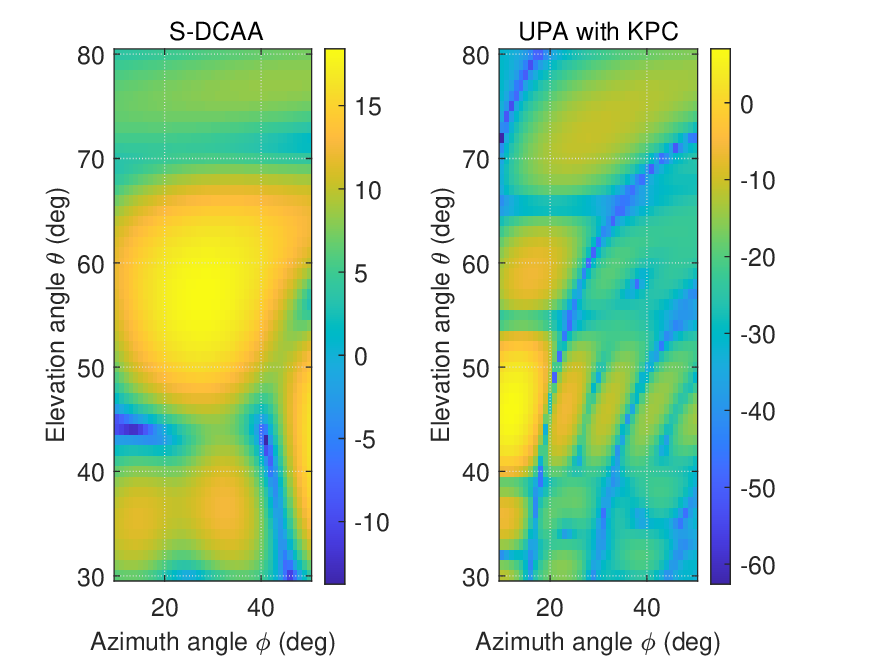}
    }
  \caption{Comparison of the aerial sensing coverage between spherical DCAA and conventional UPA for low-altitude ISAC.}\label{SDCAA_UPA_ISAC_coverage}\vspace{-10pt}
\end{figure}

Fig.~\ref{SDCAA_UPA_ISAC_coverage} compares the aerial sensing coverage achieved by the spherical DCAA and the conventional UPA benchmark.
We obtain that the spherical DCAA forms a more concentrated and higher-gain sensing beam within the prescribed coverage region, whereas the conventional UPA exhibits broader sidelobes and lower peak gains.
This result confirms that the spherical DCAA enjoys the energy-focusing capability toward the sensing region, thereby enhancing sensing performance while maintaining communication requirements.


\section{Conclusion}\label{conclusion}
In this paper, we investigated low-altitude ISAC system enabled by the spherical DCAA.
The sensing performance of spherical DCAA was characterized in terms of sensing SNR, area average probability of detection, and the CRLBs for elevation and azimuth angle estimation.
Furthermore, a low-altitude ISAC optimization problem was formulated to maximize the worst-case sensing SNR over a prescribed aerial region while satisfying the communication requirements of ground users.
To solve the resulting mixed-integer non-convex problem, a novel greedy-based joint array selection and beamforming algorithm was developed.
Simulation results demonstrated that spherical DCAA significantly outperforms conventional UPA in terms of sensing coverage, angle estimation accuracy, and communication-sensing performance tradeoff, highlighting its potential for future low-altitude ISAC systems.

\begin{appendix}[Proof of Theorem 1]\label{Appendix}
According to the chain rule, the partial derivation of $r_{n}(\phi_s,\theta_s)$ in \eqref{sensing response} w.r.t. $\phi_s$ and $\theta_s$ are given by
\begin{subequations}\label{partial deviation}
\begin{equation}
\begin{aligned}
\frac{\partial r_{n}(\phi_s,\theta_s)}{\partial\phi_s}
=
&e^{j\omega_0}\sqrt{M^2G(0,0)}
\Bigg[
\mathcal H_M'(x_{1,n})
\frac{\partial x_{1,n}}{\partial\phi_s}
\mathcal H_M(x_{2,n})
\\
&+
\mathcal H_M(x_{1,n})
\mathcal H_M'(x_{2,n})
\frac{\partial x_{2,n}}{\partial\phi_s}
\Bigg],
\end{aligned}
\label{dr_phi_theorem}
\end{equation}
\begin{equation}
\begin{aligned}
\frac{\partial r_{n}(\phi_s,\theta_s)}{\partial\theta_s}
=
&e^{j\omega_0}\sqrt{M^2G(0,0)}
\Bigg[
\mathcal H_M'(x_{1,n})
\frac{\partial x_{1,n}}{\partial\theta_s}
\mathcal H_M(x_{2,n})
\\
&+
\mathcal H_M(x_{1,n})
\mathcal H_M'(x_{2,n})
\frac{\partial x_{2,n}}{\partial\theta_s}
\Bigg].
\end{aligned}
\label{dr_theta_theorem}
\end{equation}
\end{subequations}
Note that for $\phi_s\approx \eta_{n^*}$ and $\theta_s\approx\vartheta_{n^*}$, it has
\begin{equation}\label{x_1n}
  \left\{
  \begin{aligned}
  &x_{1,n^*}\approx 0, x_{2,n^*}\approx 0\\
  &\frac{\partial x_{1,n^*}}{\partial\phi_s}\approx\cos\theta_s, \frac{\partial x_{2,n^*}}{\partial\phi_s}\approx 0\\
  &\frac{\partial x_{1,n^*}}{\partial\theta_s}\approx 0, \frac{\partial x_{2,n^*}}{\partial\theta_s}\approx 1\\
  &\mathcal{H}_M(0)=1, \mathcal{H}_M'(0) = j\pi(M-1)/2.
  \end{aligned}
  \right.
\end{equation}
For other sUPAs $n\neq n^*$, it can be assumed that $\mathcal{H}_M(x_{1,n}) $ and $\mathcal{H}_M(x_{2,n})$ are negligible, since the target direction deviates from the boresights of these sUPAs, thus we have  $\frac{\partial r_{n}(\phi_s,\theta_s)}{\partial\phi_s}\approx 0$ and $\frac{\partial r_{n}(\phi_s,\theta_s)}{\partial\theta_s}\approx 0$.

Therefore, according to \eqref{partial deviation} and \eqref{x_1n}, it has
\begin{equation}
\left\{
\begin{aligned}
&\mathbf{r}_{\phi}^H\mathbf{r}_{\phi}\approx\left|
\frac{\partial r_{n^*}(\phi_s,\theta_s)}{\partial\phi_s}
\right|^2
\approx
M^2G(0,0)
\frac{\pi^2(M-1)^2}{4}
\cos^2\theta_s\\
&\mathbf{r}_{\theta}^H\mathbf{r}_{\theta}\approx\left|
\frac{\partial r_{n^*}(\phi_s,\theta_s)}{\partial\theta_s}
\right|^2
\approx
M^2G_{n^\star}(0,0)
\frac{\pi^2(M-1)^2}{4}\\
&\mathbf{r}_{\phi}^H\mathbf{r}_{\theta}\approx \left(\frac{\partial r_{n^*}(\phi_s,\theta_s)}{\partial\phi_s}\right)^{\dagger}\frac{\partial r_{n^*}(\phi_s,\theta_s)}{\partial\theta_s}\approx 0\\
&\mathbf{r}_{\theta}^H\mathbf{r}_{\phi}\approx \left(\frac{\partial r_{n^*}(\phi_s,\theta_s)}{\partial\theta_s}\right)^{\dagger}\frac{\partial r_{n^*}(\phi_s,\theta_s)}{\partial\phi_s}\approx 0.
\end{aligned}
\right.
\end{equation}
Thus, it has
\begin{equation}\label{FIM}
\left\{
\begin{aligned}
&J_{\phi\phi}\approx\frac{
|\alpha_s|^2
\pi^2(M-1)^2
M^2
G(0,0)
\cos^2\vartheta_n
}{
2\sigma_s^2
}\\
&J_{\theta\theta} \approx \frac{
|\alpha_s|^2
\pi^2(M-1)^2
M^2
G(0,0)
}{
2\sigma_s^2
}\\
& J_{\phi\theta}=J_{\theta\phi}\approx 0.
\end{aligned}
\right.
\end{equation}
By substituting \eqref{FIM} into \eqref{CRLB}, the approximation of CRLBs for azimuth and elevation angle estimation can be derived.
This completes the proof.
\end{appendix}

\bibliographystyle{IEEEtran}
\bibliography{Spherical_DCAA_ref}

\end{document}